\documentclass[final,3p,times]{elsarticle}

\usepackage[utf8]{inputenc}
\usepackage[T1]{fontenc}
\usepackage{amsmath,amssymb}
\usepackage{booktabs}
\usepackage{multirow}
\usepackage{tabularx}
\usepackage{graphicx}
\usepackage{url}
\usepackage{xcolor}
\usepackage[colorlinks=true,
            linkcolor=blue,
            citecolor=blue,
            urlcolor=blue,
            bookmarks=false]{hyperref}

\journal{Fusion Engineering and Design}
\biboptions{sort&compress}
\begin{document}

\begin{frontmatter}

\title{MGKDB: An IMAS-aligned multicode gyrokinetic simulation database for reproducible fusion turbulence modeling and data-driven analysis}

\author[aff:sophelio,aff:uta]{Craig Michoski\corref{cor1}}
\ead{michoski@sophel.io}
\cortext[cor1]{Corresponding author.}

\author[aff:sophelio,aff:uta]{David R. Hatch}
\author[aff:sophelio]{Dongyang Kuang}
\author[aff:sophelio]{Matthew Waller}
\author[aff:ucsd]{Chris Holland}
\author[aff:differ,aff:tue,aff:rub]{M.J. Pueschel}
\author[aff:mit]{Nathan T. Howard}
\author[aff:ga]{Joseph McClenaghan}
\author[aff:ga]{Tom F. Neiser}
\author[aff:ga]{Max T. Curie}
\author[aff:sophelio]{Venkitesh Ayyar}
\author[aff:uta]{Joseph Schmidt}
\author[aff:uta]{Leonhard A. Leppin}

\author[aff:mit]{Aaron Ho}

\author[aff:sophelio]{Tapan Ganatma Nakkina}

\author[aff:ukaea]{Bhavin S. Patel}

\author[aff:piim]{Yann Camenen}

\address[aff:sophelio]{Sophelio LLC, Austin, Texas, USA}

\address[aff:differ]{Dutch Institute for Fundamental Energy Research,
5612 AJ Eindhoven, The Netherlands}

\address[aff:tue]{Eindhoven University of Technology,
5600 MB Eindhoven, The Netherlands}

\address[aff:rub]{Department of Physics \& Astronomy,
Ruhr-Universit{\"a}t Bochum,
D-44780 Bochum, Germany}

\address[aff:ga]{General Atomics, San Diego, California, USA}

\address[aff:ucsd]{University of California San Diego,
La Jolla, California, USA}

\address[aff:uta]{The University of Texas at Austin,
Austin, Texas, USA}

\address[aff:mit]{Plasma Science and Fusion Center,
Massachusetts Institute of Technology,
Cambridge, Massachusetts, USA}

\address[aff:piim]{Aix Marseille Univ, CNRS, PIIM,
Marseille, France}

\address[aff:ukaea]{United Kingdom Atomic Energy Authority,
Abingdon, United Kingdom}

\begin{abstract}
Expensive fusion simulations are commonly preserved in code-specific formats
that limit discovery, comparison, and reuse. We present the Multiscale GyroKinetic DataBase (MGKDB), an open-source
software framework and curated archive that converts heterogeneous simulation
campaigns into traceable scientific records. Each record links code-native
inputs and outputs to provenance and quality metadata, an IMAS-aligned physics
representation, and derived diagnostics, preserving model-specific evidence
while enabling common-field queries. Production pathways support linear and
nonlinear GENE and CGYRO calculations and reduced quasilinear TGLF evaluations.
At the September~1, 2026 snapshot, MGKDB contained \(1{,}068{,}089\) records,
nearly all of which included a populated gyrokinetics IMAS branch. The software is
openly available, while access to the NERSC-hosted production records is
managed. Three demonstrations show how these linked representations support scientific
reuse. Standardized quantities stored in the Diagnostics branch enable
population-scale analysis of archived linear modes; common input coordinates
reveal coverage, redundancy, and campaign-driven sampling structure across a
multicode collection; and record-level retrieval of native CGYRO inputs drives
matched TGLF calculations and produces a traceable dataset for exploratory
surrogate modeling. Together, these examples demonstrate how MGKDB supports
archive characterization, candidate cross-code and cross-fidelity
comparisons, campaign planning, and reproducible data-driven modeling without
treating different models as automatically equivalent.
\end{abstract}



\end{frontmatter}

\section{Introduction}
\label{sec:introduction}

Predictive modeling of magnetically confined fusion plasmas increasingly depends on a hierarchy of computational tools for describing microinstabilities, turbulence, and transport. Gyrokinetic theory makes high-fidelity kinetic calculations tractable by removing the fast gyromotion while retaining the dynamics needed to describe turbulence in strongly magnetized plasmas~\cite{gyro1,gyro2}. Modern gyrokinetic solvers such as GENE and CGYRO provide detailed linear and nonlinear calculations~\cite{gene,cgyro}, while reduced models such as TGLF approximate selected gyrokinetic physics to support faster transport prediction~\cite{tglf}. Nevertheless, nonlinear gyrokinetic calculations remain sufficiently expensive that individual runs and simulation campaigns represent substantial scientific and computational investments. Their value can extend well beyond the study for which they were generated, e.g. archived calculations can support code verification and validation, cross-code and cross-fidelity benchmarking, reduced-model development, uncertainty studies, machine-learning surrogates, and the design of subsequent simulation campaigns~\cite{vandeplassche2020,rodriguezfernandez2024portals}. In practice, however, many results remain accessible only as code-specific directory trees, private campaign archives, or reduced tables that omit information needed for later interpretation and reproduction.

Reusing a plasma simulation requires more than retaining its principal output arrays. The associated geometry and equilibrium, species definitions, normalization conventions, numerical resolution, solver settings, code version, convergence information, and provenance must remain connected to the result. These requirements are particularly important when combining calculations from models at different levels of fidelity, whose physical assumptions, input conventions, coordinate systems, file formats, and available diagnostics can differ substantially and may change as the software evolves. Without a common representation, cross-code and cross-fidelity analyses require bespoke conversion and normalization work. Dataset construction for machine learning repeatedly reproduces the same extraction and cleaning steps, and archive-scale analyses can unknowingly inherit duplicated samples, incomplete metadata, or strong biases toward particular campaigns. The underlying infrastructure problem is therefore to connect the complete code-native artifacts needed for reproduction with standardized, searchable representations of their physics content, without discarding model-specific information.

Several elements of this infrastructure already exist within the fusion community. MDSplus provides durable, queryable storage for experimental and analysis data~\cite{mdsplus,mdsplus2}, with recent extensions addressing the access patterns required by machine-learning workflows~\cite{mdsplusml}. The ITER Integrated Modelling and Analysis Suite (IMAS) defines
machine-independent Interface Data Structures for exchanging experimental and
simulated fusion data~\cite{Imbeaux_2015}. Its dedicated gyrokinetics IDS
establishes common conventions and normalizations for flux-tube gyrokinetic
inputs and outputs and is intended to facilitate multicode benchmarking, data
exchange, and database construction~\cite{camenen2026gyrokineticsids}. The
release of the IMAS infrastructure under open-source licenses further
strengthens its role as a common foundation for fusion modeling and
analysis~\cite{iter_imas_open_2025}. For gyrokinetics specifically, \textit{pyrokinetics} supplies code-aware readers, writers, and normalization conversions through a common Python interface~\cite{patel2024pyrokinetics}, while the Gyro-Kinetic Database project has pursued the complementary objective of organizing gyrokinetic results for shared analysis~\cite{gkdb}. More broadly, computational-materials platforms demonstrate how standardized simulation collections can become reusable scientific infrastructure when numerical data, provenance, and programmatic interfaces are developed together~\cite{materialscloud,materialsproject,oqmd}. The remaining operational challenge is to integrate these principles into a persistent multicode and multimodel fusion-simulation archive that retains reproducible native files while exposing standardized physics quantities and diagnostics for database-scale querying.

The Multiscale GyroKinetic DataBase (MGKDB) addresses this challenge by
combining an open-source software framework with a curated production database
hosted at the National Energy Research Scientific Computing Center (NERSC).
Its present implementation is centered on gyrokinetic simulations and
associated reduced transport models.  The current focus is on GENE, CGYRO, and TGLF, although the \emph{Pyrokinetics} interface should make additional gyrokinetic codes easily portable, including GS2, GX, GKW, and STELLA.  Despite the current focus on gyrokinetics, the underlying document architecture
is intended to accommodate additional classes of fusion simulation, including
reduced-fluid and magnetohydrodynamic models, as suitable standardized
representations and ingestion tools are developed. Each simulation record
links four complementary forms of information: code-native input and output
files, provenance and quality metadata, standardized physics data, and
precomputed diagnostics. For the presently supported gyrokinetic and
reduced-model records, the standardized physics content is stored in a
gyrokinetics IMAS representation. The native files preserve the information
required to inspect or reproduce a calculation, while the standardized branch
allows quantities with compatible physical definitions to be queried across
codes through common field names, data structures, and normalization
conventions.

This IMAS alignment also connects MGKDB naturally to the broader ecosystem of
IMAS-compatible databases, data platforms, integrated-modeling environments,
and analysis workflows. Rather than requiring a new project-specific
translation for every connection, compatible data can be exchanged through
shared IMAS Interface Data Structures and field definitions. IMAS compatibility
does not by itself guarantee that quantities produced by different models are
physically interchangeable, but it provides a common structural and semantic
layer through which appropriately matched records can be discovered,
exchanged, and incorporated into downstream workflows. The document-based
schema can evolve as new model classes, codes, diagnostics, and IMAS fields are
introduced, and the open-source software and interfaces can be adapted as a
foundation for databases organized around other IMAS data structures.

Upload,
conversion, retrieval, and analysis tools are distributed openly; access to
the curated production instance presently requires NERSC and MGKDB
credentials. Thus, ``open source'' refers to the software, schema, and
interfaces, whereas access to the hosted scientific records is managed under
explicitly stated conditions, consistent with the FAIR principles of
findability, accessibility, interoperability, and
reusability~\cite{fair}.

At the dated reference snapshot of September~1, 2026, MGKDB contained
$1{,}068{,}089$ records from GENE, CGYRO, and TGLF, of which
$1{,}067{,}980$ ($99.99\%$) included a populated gyrokinetics IMAS branch. This dated count is a reproducible reference point rather than a fixed endpoint; i.e., MGKDB is an actively maintained and growing production archive, with new simulation results currently being added on a roughly weekly basis as
campaigns are completed and eligible records are ingested.  In this paper, we evaluate the practical utility of this infrastructure through three complementary demonstrations. First, standardized output descriptors from $6{,}053$ linear GENE simulations are used to characterize the instability class and neighborhood consistency based on a physics-motivated mode-fingerprint representation. Second, the distribution of $24{,}563$ linear GENE and CGYRO entries is examined in a shared 13-dimensional control-parameter space to identify clustering, redundancy, and sparsely sampled directions relevant to future campaign design. Third, archived nonlinear CGYRO operating points are used to generate TGLF evaluations at the same operating points and assembled into a cross-fidelity dataset for exploratory surrogate modeling. These examples test different stages of the scientific-data lifecycle: population-scale analysis of standardized outputs, assessment of archive coverage from standardized inputs, and construction of a reproducible downstream modeling workflow.

The principal contribution of MGKDB is therefore not any individual
statistical model or visualization, but an extensible and actively growing
infrastructure through which multicode and multimodel fusion simulations can
be treated as traceable, queryable scientific populations and connected to the
broader ecosystem of IMAS-compatible databases and workflows. The remainder of this paper introduces the relevant physical and data-standardization context, describes the MGKDB architecture and record schema, documents its access and user interfaces, presents the three database-enabled use cases, and discusses schema evolution, external workflow integration, and priorities for continued community development.

\section{Fusion-model hierarchy and data-standardization layers}
\label{sec:model_context}

\begin{figure}[htb!]
\centering
\includegraphics[width=0.99\linewidth]
{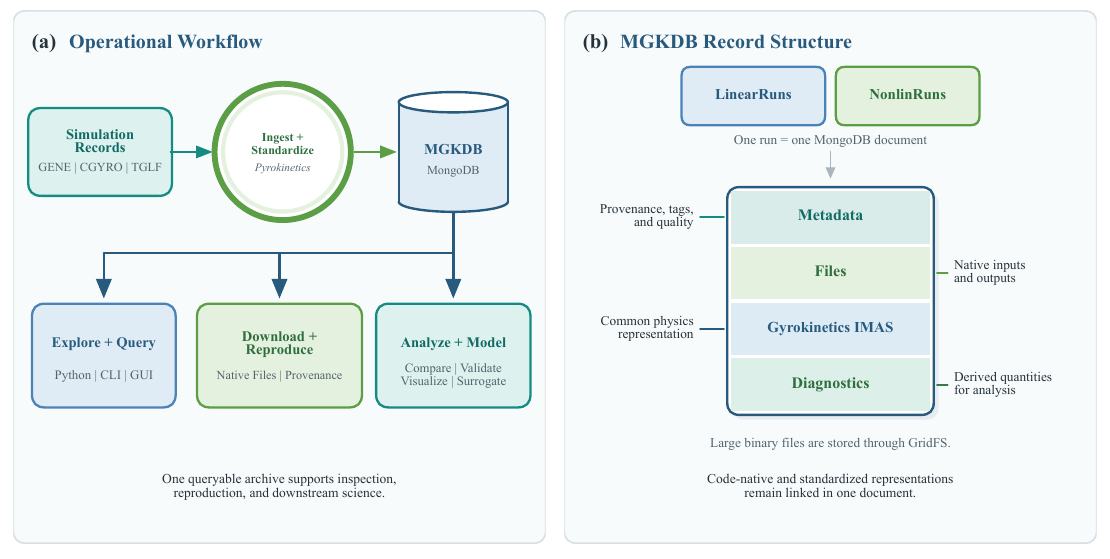}
\caption{MGKDB operational workflow and record architecture.
(a) Code-native GENE and CGYRO calculations and TGLF reduced-model
evaluations are ingested through code-aware readers and standardized
using \textit{Pyrokinetics}. The resulting records can be queried and
retrieved through command-line, Python, and graphical interfaces for
reproducibility, same-fidelity cross-code comparison, cross-fidelity
analysis, archive characterization, and data-driven modeling.
(b) Each calculation is stored as a MongoDB document containing four
principal branches: provenance and quality metadata, code-native files,
an IMAS-aligned standardized physics representation, and derived
diagnostics. The separation between native and standardized content
preserves model-specific information while enabling common queries
across supported codes and model classes.}
\label{fig:mgkdb_architecture}
\end{figure}

MGKDB separates four related but distinct layers of a reusable simulation record: the physical model that produces the calculation, the code-native files needed to inspect or reproduce it, the translation software used to interpret those files, and the standardized representation used for database-scale querying. This separation makes heterogeneous calculations discoverable and comparable without implying that different physical models or numerical implementations are equivalent. In the current implementation, GENE and CGYRO provide high-fidelity gyrokinetic calculations, TGLF provides reduced quasilinear transport evaluations, \textit{Pyrokinetics} performs code-aware translation and normalization, and an IMAS-aligned gyrokinetics representation supplies the common physics structure stored by MGKDB. The relationship among these physical-model, translation, standardized-data, and archival layers is summarized in Fig.~\ref{fig:mgkdb_architecture}.

\subsection{Present model hierarchy}
\label{sec:model_hierarchy}

The kinetic evolution of a magnetized plasma may be described by the Vlasov--Maxwell system, or by a Vlasov--Fokker--Planck formulation when collisional effects are retained. Gyrokinetic theory exploits the separation between the rapid gyromotion and the lower-frequency dynamics of interest, averaging over the gyrophase while retaining the kinetic behavior relevant to microinstabilities and turbulent transport~\cite{gyro1,gyro2}. The resulting distribution function is represented in a reduced five-dimensional phase space for each kinetic species rather than in the full six-dimensional particle phase space. This reduction makes first-principles turbulence calculations more tractable, although nonlinear gyrokinetic simulations can still require substantial computational resources.

GENE~\cite{gene} and CGYRO~\cite{cgyro} are Eulerian gyrokinetic solvers that evolve gyro-averaged distribution functions for the kinetic species represented in a calculation. Depending on the selected configuration, they can include electromagnetic fluctuations, collisions, multiple species, and realistic magnetic geometry. Both codes support linear calculations, which characterize instability growth rates, real frequencies, and mode structures, as well as nonlinear calculations, which evolve the saturated turbulent state and produce the fluxes required for transport prediction.

TGLF~\cite{tglf} occupies a different level of this modeling hierarchy. It is a reduced quasilinear transport model informed by gyrokinetic physics, not a gyrokinetic solver. TGLF approximates the relevant linear response using a trapped gyro-Landau-fluid formulation and combines those results with saturation rules to estimate turbulent fluxes at substantially lower computational cost than a corresponding nonlinear gyrokinetic calculation. Its inclusion in MGKDB enables workflows that connect high-fidelity simulations to reduced-model evaluations at matched operating points.

\begin{table}[htb!]
\centering
\small
\caption{Physical roles of the simulation tools presently supported by
MGKDB. The classification distinguishes the underlying model from its
operational use in the database.}
\label{tab:model_hierarchy}

\begingroup
\setlength{\tabcolsep}{5pt}
\renewcommand{\arraystretch}{1.18}

\begin{tabularx}{\linewidth}{@{}
    >{\raggedright\arraybackslash}l
    >{\raggedright\arraybackslash}p{0.28\linewidth}
    >{\raggedright\arraybackslash}X@{}}
\toprule
\textbf{Tools} &
\textbf{Model class} &
\textbf{Present role in MGKDB} \\
\midrule

\textbf{CGYRO, GENE} &
Eulerian gyrokinetic solvers &
Linear instability calculations and nonlinear turbulent-transport
simulations. \\

\textbf{TGLF} &
Reduced quasilinear transport model &
Low-cost transport and instability evaluations used in reduced-model and
cross-fidelity workflows. \\

\bottomrule
\end{tabularx}

\endgroup
\end{table}

We use \emph{cross-code comparison} to refer to comparisons between independent implementations at comparable levels of physical fidelity, such as GENE and CGYRO calculations constructed with commensurate assumptions. We use \emph{cross-fidelity comparison} for relationships between models at different levels of approximation, such as nonlinear CGYRO simulations and TGLF evaluations. Standardized storage facilitates both forms of analysis, but it does not establish that two records constitute a valid comparison. Compatibility of geometry, species content, physical assumptions, numerical resolution, and output definitions must still be assessed.

The current production database stores linear gyrokinetic records in the \textbf{LinearRuns} collection and stores nonlinear gyrokinetic and quasilinear TGLF records in \textbf{NonlinRuns}. These names reflect the historical and operational development of the archive and should not be interpreted as a complete taxonomy of fusion models. The underlying document pattern is not intrinsically restricted to gyrokinetics. Reduced-fluid and magnetohydrodynamic frameworks, such as BOUT++~\cite{dudson2009bout}, could be incorporated through appropriate code-aware readers and standardized physics branches. Such support is an architectural direction rather than a capability demonstrated in the present work: the production ingestion and standardization pathways reported here are currently limited to GENE, CGYRO, and TGLF.

\subsection{IMAS as the common representation}
\label{sec:imas_context}

The Integrated Modelling and Analysis Suite (IMAS) provides a machine-independent framework for representing experimental and simulated fusion data~\cite{Imbeaux_2015}. Its Data Dictionary defines a hierarchy of Interface Data Structures (IDSs), with standardized field names, organization, units, and associated metadata. IMAS therefore specifies how supported physical quantities are represented; it does not prescribe which physical model must generate them or require that all model-specific information be discarded.

MGKDB presently uses an IMAS-aligned representation of the gyrokinetics IDS as its standardized physics branch. This branch records common quantities such as normalizing values, model descriptors, flux-surface geometry, species properties, collisionality, code information, and available linear or nonlinear outputs. Representing these quantities under common paths allows queries to be formulated in terms of physical content rather than the native syntax of a particular solver. For example, a query for magnetic shear, normalized gradients, growth rates, or turbulent fluxes can address a consistent database path even when the corresponding values originated in different code-specific files.

The standardized branch complements rather than replaces the native simulation files. Some solver settings, diagnostics, intermediate data, or model-specific quantities may not have direct counterparts in the current gyrokinetics IDS. MGKDB therefore retains the original inputs and outputs alongside the standardized representation. The native files remain available for detailed inspection and reproduction, whereas the standardized branch supplies the common subset needed for discovery, filtering, and population-scale analysis.

Common field names and normalization conventions also do not make two calculations automatically interchangeable. A scientifically meaningful comparison must still account for model fidelity, geometry, species treatment, collisions, electromagnetic assumptions, resolution, convergence, and other information stored in the record. IMAS standardization makes such comparisons more systematic and auditable; it does not replace the physical judgment required to construct them.

The gyrokinetics IDS is appropriate for the model classes currently represented in MGKDB because GENE, CGYRO, and the relevant TGLF quantities share a substantial set of inputs and transport-related outputs. Incorporating reduced-fluid, extended-MHD, or other fusion models may require different IMAS IDSs or additional standardized branches. The MongoDB document architecture permits such branches to coexist with the present gyrokinetics representation, but their schemas and validation requirements must be defined as each new model family is added.

\subsection{Pyrokinetics as the translation layer}
\label{sec:pyrokinetics_context}

The standardized gyrokinetics representation in MGKDB is generated using the open-source \textit{Pyrokinetics} package~\cite{patel2024pyrokinetics}. Pyrokinetics\footnote{Documentation: \url{https://pyrokinetics.readthedocs.io/en/latest/}; source: \url{https://github.com/pyro-kinetics/pyrokinetics}.} provides code-aware readers, writers, normalization conversions, and common data objects for supported gyrokinetic and reduced transport codes, including GENE, CGYRO, and TGLF.  Within the MGKDB ingestion workflow, Pyrokinetics acts as the translation layer between code-native artifacts and the standardized database branch. It parses supported inputs and outputs, interprets the normalization conventions used by the source code, converts supported quantities to the common representation, and populates the gyrokinetics IDS structure stored in the database. The original files are retained in parallel, so the standardized representation does not become the sole record of the calculation.

This dual representation addresses two different requirements. Code-native files preserve model-specific detail and provide the strongest basis for reproducing a run. The standardized branch provides the stable physical interface required for queries, cross-code analysis, archive diagnostics, and downstream dataset construction. Because the code-specific parsing logic resides in Pyrokinetics rather than in the MongoDB schema, improvements to readers, writers, and normalization handling can be incorporated without redesigning the overall MGKDB document model.

Pyrokinetics also supports post-processing and visualization outside the database, but its principal architectural role in MGKDB is controlled translation rather than the elimination of all differences among models. Not every native quantity has a one-to-one standardized counterpart, and the validity of a cross-code or cross-fidelity comparison continues to depend on the stored provenance and model description. Changes to the IMAS schema and the corresponding migration of existing records are discussed further in Sec.~\ref{sec:evolution}.

Together, these layers define both the present scope and the intended extensibility of MGKDB. The current production system demonstrates a shared archive for two gyrokinetic solvers and one reduced quasilinear model using a common gyrokinetics representation. Its broader contribution is an architecture in which additional fusion-model classes can be incorporated without erasing the distinctions among their governing equations, approximations, native data products, or domains of validity.

\section{MGKDB system overview and design objectives}
\label{sec:overview}

MGKDB is an end-to-end scientific-data system rather than only a database instance. It comprises two connected components: an open-source software framework\footnote{\url{https://github.com/Sophelio/MGKDB}} for ingesting, standardizing, querying, retrieving, and analyzing fusion-simulation records, and a shared production archive hosted at NERSC. The software can be used independently to create local or institutionally managed databases, whereas the NERSC deployment provides a curated environment through which participating researchers can contribute, discover, and reuse simulation records under a common data model. The current production pathways support the gyrokinetic solvers GENE and CGYRO and the reduced quasilinear transport model TGLF.

The basic unit of MGKDB is not an isolated output array or directory, but a traceable simulation record. As described in Sec.~\ref{sec:model_context}, each record links the code-native inputs and outputs to provenance and quality metadata, an IMAS-aligned standardized physics representation, and derived diagnostics. Retaining these elements together serves two complementary purposes. Native artifacts preserve model-specific information and provide the strongest available basis for inspection and reproduction, while standardized quantities allow physically relevant subsets of the archive to be discovered and analyzed without first decoding every source format independently. The record therefore remains specific enough to preserve the meaning of the original calculation while becoming structured enough to participate in population-scale queries and downstream workflows.

MGKDB is organized around four primary design objectives:

\begin{enumerate}
\item \textbf{Reproduction and auditability} ---
Preserve code-native inputs and outputs together with provenance,
model configuration, code-version information, and quality metadata,
allowing calculations to be inspected and, where the required software,
dependencies, and permissions remain available, reproduced.

\item \textbf{Interoperable discovery and comparison} ---
Represent commonly used physical quantities under consistent database
paths so that potentially compatible records can be located across codes
and model fidelities. Standardization identifies candidate comparisons;
scientific compatibility must still be established from the stored model
assumptions, numerical configuration, and provenance.

\item \textbf{Reuse in downstream workflows} ---
Support programmatic construction of analysis-ready datasets for
visualization, quality assessment, same-fidelity cross-code studies,
cross-fidelity analysis, reduced-model evaluation, surrogate modeling,
and other data-driven applications.

\item \textbf{Archive stewardship and campaign planning} ---
Characterize the population of stored calculations to identify
duplication, sampling biases, and sparsely explored regions, allowing
accumulated simulation results to inform the design of subsequent
computational campaigns.

\end{enumerate}

\begin{figure}[htb!]
\centering
\includegraphics[width=0.95\linewidth]
{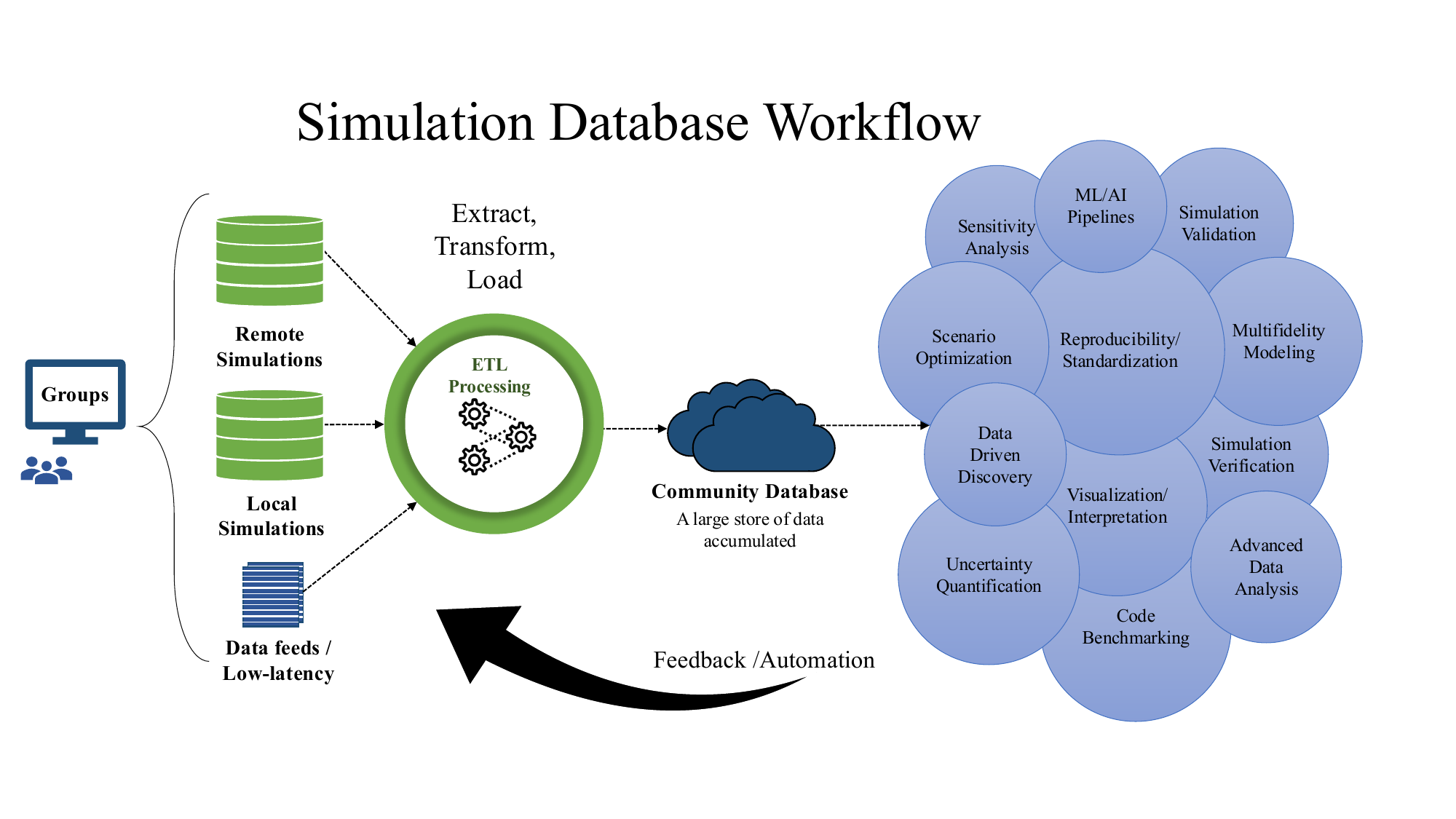}
\caption{Database-centered scientific lifecycle implemented by MGKDB.
Code- and model-specific inputs and outputs are ingested through an
extract--transform--load (ETL) process that preserves provenance and
native artifacts while mapping supported quantities into standardized
representations. The resulting queryable records support inspection and
reproduction, visualization, same-fidelity cross-code comparison,
cross-fidelity analysis, archive characterization, validation workflows,
and data-driven modeling. Findings from these activities can then guide
subsequent simulation campaigns. MGKDB currently implements this
lifecycle for GENE and CGYRO gyrokinetic calculations and TGLF
reduced-model evaluations, with an architecture intended to accommodate
additional fusion-model classes as suitable ingestion and
standardization pathways are developed.}
\label{fig:generic}
\end{figure}

Together, these objectives define a reusable scientific-data lifecycle in
which completed simulations become inputs to new analysis and modeling
workflows, and the resulting findings can guide subsequent computational
campaigns (Fig.~\ref{fig:generic}).

The software framework and the hosted archive have different access models. The source code, database schema, and user documentation are openly available, and the framework can be deployed independently of the production service. Access to the shared NERSC archive presently requires both NERSC authorization and MGKDB credentials, as described in Sec.~\ref{sec:data_schema}. The hosted archive therefore operates as a shared, curated service rather than as a public bulk-download repository. This distinction is compatible with the FAIR principle that access conditions should be explicit and machine-actionable, but need not be unrestricted~\cite{fair}. It also recognizes that the software and the scientific records are distinct research objects with different requirements for discovery, access, interoperability, and reuse~\cite{fair4rs}.

The present scope of the production system should likewise be distinguished from the extensibility of its architecture. GENE, CGYRO, and TGLF are supported because code-aware readers, normalization mappings, standardized physics fields, and ingestion procedures have been implemented for them. The document structure can accommodate additional fusion-model classes, including reduced-fluid and magnetohydrodynamic models, but architectural compatibility alone does not constitute production support. Each new model family requires appropriate native-file readers, model-specific provenance, standardized data mappings, validation checks, and user-facing documentation before its records can be interpreted consistently.

MGKDB facilitates scientific comparison but does not replace the scientific criteria required to validate one calculation against another. A shared field name does not guarantee agreement in geometry, species content, physical assumptions, normalization, resolution, convergence, or domain of validity. The database instead makes these considerations more explicit and auditable by keeping standardized quantities linked to their native artifacts and provenance. In the same way, the presence of executable workflows does not make every archived record suitable for machine learning: cohort definition, missingness, duplication, campaign structure, and validation design remain analysis-level decisions.

The following sections evaluate the system at progressively more concrete levels. Section~\ref{sec:data_schema} documents the production deployment, dated archive census, storage architecture, and record schema. Section~\ref{sec:interfaces} describes the command-line, Python, and graphical interfaces through which users contribute, query, and retrieve records. Section~\ref{sec:case_study} then exercises the lifecycle of Fig.~\ref{fig:generic} through population-scale analysis of standardized outputs, assessment of parameter-space coverage, and automated construction of a cross-fidelity modeling dataset. Together, these sections test whether the architecture described above can convert heterogeneous campaign artifacts into traceable, reusable scientific resources.

\section{Production archive, storage architecture, and record schema}
\label{sec:data_schema}

This section describes the production MGKDB instance at a fixed reference
date, the storage choices that support heterogeneous simulation records, and
the scientific meaning of the principal document branches. Operational
instructions for authentication, upload, query, and retrieval are deferred to
Sec.~\ref{sec:interfaces}; the focus here is the structure required to
interpret and reuse an archived calculation.

\subsection{Production deployment and dated census}

The production MGKDB instance is hosted at NERSC\footnote{\url{https://www.nersc.gov}} and
is accessed within the NERSC computing environment. Unless otherwise stated,
all archive-wide counts in this paper refer to a direct query of the
production instance on September~1, 2026. The scientific analyses in
Sec.~\ref{sec:case_study} use separate, fixed extracts; their sizes therefore
should not be interpreted as contemporaneous production totals.

At the reference snapshot, MGKDB contained $1{,}068{,}089$ records. The
\textbf{LinearRuns} collection comprised $37{,}287$ linear GENE and CGYRO
calculations, or $3.49\%$ of the archive. The \textbf{NonlinRuns} collection
contained $1{,}030{,}802$ records, comprising nonlinear GENE and CGYRO
calculations and quasilinear TGLF evaluations. A populated gyrokinetics IMAS
branch was present in $1{,}067{,}980$ records ($99.99\%$). The remaining
$109$ records comprise $108$ legacy nonlinear GENE entries with an
unpopulated legacy gyrokinetics representation and one nonlinear GENE record
lacking the branch. Table~\ref{tab:snapshot} separates this production census from the fixed cohorts used in the three demonstrations.

\begin{table}[tbh]
\centering
\small
\caption{MGKDB production census and analysis cohorts. Production counts refer
to the September~1, 2026 snapshot. The analysis cohorts are fixed extracts
used for the indicated results and are not subsets drawn from that same
snapshot.}
\label{tab:snapshot}
\begin{tabularx}{\linewidth}{@{}
    >{\raggedright\arraybackslash}X
    >{\raggedright\arraybackslash}p{0.23\linewidth}
    >{\raggedleft\arraybackslash}p{0.14\linewidth}@{}}
\toprule
\textbf{Quantity} & \textbf{Snapshot or extract} & \textbf{Count} \\
\midrule
\multicolumn{3}{@{}l}{\textbf{Production census}} \\
All records & September~1, 2026 & $1{,}068{,}089$ \\
\quad \textbf{LinearRuns} total & September~1, 2026 & $37{,}287$ \\
\qquad Linear GENE & September~1, 2026 & $17{,}848$ \\
\qquad Linear CGYRO & September~1, 2026 & $19{,}439$ \\
\quad \textbf{NonlinRuns} total & September~1, 2026 & $1{,}030{,}802$ \\
\qquad Nonlinear GENE & September~1, 2026 & $292$ \\
\qquad Nonlinear CGYRO & September~1, 2026 & $840$ \\
\qquad TGLF (quasilinear) & September~1, 2026 & $1{,}029{,}670$ \\
Records with a populated gyrokinetics IMAS branch &
September~1, 2026 & $1{,}067{,}980$ \\
\midrule
\multicolumn{3}{@{}l}{\textbf{Fixed analysis cohorts}} \\
Fingerprint analysis, Sec.~\ref{sec:contents} & July 2026 & $6{,}053$ \\
Coverage analysis, Sec.~\ref{sec:coverage} & April~26, 2026 & $24{,}563$ \\
Surrogate demonstration, Sec.~\ref{sec:surrogates} & Campaign extract & $50$ \\
\bottomrule
\end{tabularx}
\end{table}

The deployment metrics associated with the production snapshot report
approximately $360.2$~GB stored on disk and a logical data volume of
approximately $457.4$~GB before storage-engine compression. These values are
operational estimates for the deployment rather than sums of user-visible
file sizes. The deployment includes database documents and associated GridFS content,
including the retained code-native artifacts linked to individual simulation
records. The
dated census and the corresponding analysis provenance are summarized further
in Supplementary Sec.~\ref{sec:supp:snapshots}.

\subsection{Document and binary-file storage}

MGKDB uses MongoDB~\cite{banker2012mongodb,chodorow2010mongodb} because a
simulation run maps naturally to a nested scientific document: provenance,
model settings, species and geometry data, outputs, and optional code-specific
fields can remain linked under one record identifier. The document model also
allows model-specific fields to coexist without forcing every record to
contain an identical set of columns. This flexibility is important for an
archive that contains several codes, multiple fidelity levels, and records
created under successive schema versions. It does not remove the need for
controlled field definitions, migration, and validation; those responsibilities
are implemented in the ingestion software and standardized data layer.

MongoDB limits an individual BSON document to $16$~MiB.\footnote{MongoDB,
``MongoDB Limits and Thresholds,''
\url{https://www.mongodb.com/docs/manual/reference/limits/} (accessed
August~12, 2026).} Native simulation files that exceed, or are unsuitable for,
that document representation are stored through GridFS. GridFS divides a file
into linked chunks and records the associated file metadata and binary content
in paired collections, conventionally \texttt{fs.files} and
\texttt{fs.chunks}.\footnote{MongoDB, ``GridFS,''
\url{https://www.mongodb.com/docs/manual/core/gridfs/} (accessed August~12,
2026).} The corresponding identifiers are linked to the simulation record, so
large code-native artifacts remain discoverable through the same run-level
query as the structured metadata and physics quantities.

For supported gyrokinetic and reduced-model records, MGKDB complements this
document-level flexibility with an IMAS-aligned gyrokinetics representation.
IMAS supplies the shared scientific organization, while the Ordered
Multidimensional Array Structures (OMAS) library provides a Python
representation compatible with the IMAS data model and IMAS-aware integrated
workflows~\cite{Imbeaux_2015,lyons2023step}.\footnote{OMAS documentation:
\url{https://gafusion.github.io/omas/} (accessed August~12, 2026).} This layer
supports queries formulated in terms of physical quantities rather than native
file syntax. It does not convert MGKDB into a relational database, nor does it
make calculations with different model assumptions automatically equivalent.

\subsection{Scientific record schema}
\label{sec:schema}

Each primary MongoDB document represents one simulation run. The production
archive uses two top-level collections: \textbf{LinearRuns} contains linear
GENE and CGYRO calculations, whereas \textbf{NonlinRuns} contains nonlinear
GENE and CGYRO calculations together with quasilinear TGLF evaluations. The
collection names are therefore historical operational partitions, not a
complete physical taxonomy. The model family and run class should be inferred
from explicit metadata fields such as \texttt{sim\_type},
\texttt{IsLinear}, and \texttt{quasi\_linear}. The physical organization of a run document is shown in
Fig.~\ref{fig:mgkdb_architecture}(b). Table~\ref{tab:record_branches}
summarizes the role of each principal branch. Detailed field and file
inventories are provided in Supplementary
Sec.~\ref{sec:supp:schema_reference}.

\begin{table}[tbh]
\centering
\small
\caption{Principal branches of an MGKDB simulation record. Availability can
vary for legacy records and for code-specific optional outputs.}
\label{tab:record_branches}
\begin{tabularx}{\linewidth}{@{}
    >{\raggedright\arraybackslash}p{0.18\linewidth}
    >{\raggedright\arraybackslash}p{0.30\linewidth}
    >{\raggedright\arraybackslash}X@{}}
\toprule
\textbf{Branch} & \textbf{Contents} & \textbf{Scientific role} \\
\midrule
\texttt{Meta} & Provenance, identifiers, model configuration, timestamps, and uploader-assigned review status & Establishes origin and context and supports discovery, grouping, and lineage queries \\
\texttt{Files} & Retained code-native inputs and outputs, stored directly or referenced through GridFS & Preserves the authoritative archived artifacts for inspection and, where the retained payload and software permit, reproduction \\
\texttt{gyrokinetics} & IMAS-aligned quantities translated from supported native inputs and outputs & Provides the standardized physics layer for common-path queries and candidate cross-code or cross-fidelity comparisons \\
\texttt{Diagnostics} & Preprocessed and derived arrays; currently populated for GENE records & Accelerates recurring analyses without replacing the native artifacts from which the quantities were derived \\
\texttt{Plots} & Optional stored visualizations & Provides convenient previews; these products are non-authoritative \\
\bottomrule
\end{tabularx}
\end{table}

Earlier MGKDB records used conversion scripts adapted from the open-source
Gyro-Kinetic Database project~\cite{gkdb}. Current uploads use
\textit{Pyrokinetics}~\cite{patel2024pyrokinetics} as the code-aware
translation layer. The coexistence of legacy and current records is why schema
versioning, migration, and explicit treatment of missing fields are necessary;
the migration strategy is discussed in Sec.~\ref{sec:evolution}.

\subsubsection{Provenance and quality metadata}

The \texttt{Meta} branch records who uploaded a run, how it was grouped, which
code and workflow produced it, when it was executed and ingested, and whether
it is linked to a restart or related calculation. In particular,
\texttt{user} identifies the uploader and should not, without additional
evidence, be interpreted as the sole simulation author or operator.
\texttt{run\_collection\_name} is derived from the upload-folder path and is a
practical grouping field rather than a globally persistent campaign
identifier. The combination of this field with \texttt{run\_suffix} supports
the retrieval of scans and related cases, while fields such as
\texttt{linked\_objectID} and \texttt{IsRestart} preserve explicit run
lineage when supplied.

The numerical \texttt{confidence} field is an uploader-assigned review-status
indicator, not a statistical confidence, probability of correctness, or
certificate of physical validation. Level~0 means that verification has not
been assessed, not that a run has failed. Higher levels record progressively
stronger checks, but a convergence check for one reported quantity does not
validate every output or establish agreement with experiment or another code.
This terminology follows the broader distinction between numerical
verification and physical validation~\cite{oberkampf2004verification}. The
field definitions and level-specific guidance are given in Supplementary
Tables~\ref{tab:supp:metadata_provenance}--\ref{tab:supp:confidence}.

Metadata completeness varies across legacy records. Queries that depend on an
optional field must therefore test for its presence rather than treating
absence as a negative value. For future ingestion, stable campaign
identifiers, software and translation versions, and checksums for retained
artifacts are particularly valuable because they make lineage and migration
auditable independently of folder names.

\subsubsection{Code-native artifacts}

The \texttt{Files} branch retains supported code-native inputs and outputs and
may contain additional artifacts contributed with a run. The exact payload
depends on the source code, simulation type, upload configuration, and age of
the record. Consequently, MGKDB should not be described as guaranteeing that
every historical entry contains every file required for bitwise reproduction.
Rather, the branch preserves the strongest available code-native record and
supports rerunning or auditing a calculation when the required inputs,
executable or source version, dependencies, and optional outputs are all
available. Representative retained artifacts are listed in Supplementary
Table~\ref{tab:supp:native_files}.

Within the record hierarchy, retained native files are the authoritative
archived artifacts. The standardized branch, cached diagnostics, and plots are
translated or derived products and should remain traceable to those artifacts.
This distinction allows the common schema to evolve without erasing the
original representation of the calculation.

\subsubsection{Standardized physics and derived diagnostics}

The \texttt{gyrokinetics} branch is populated by translating supported native
inputs and outputs with \textit{Pyrokinetics}; it is not computed solely from
raw output files. Its IMAS-aligned groups describe normalizing quantities,
model assumptions, flux-surface geometry, species, collisions, code
information, and available linear or nonlinear results. Common field paths and
normalizations allow users to discover potentially comparable records and to
construct code-independent analysis tables. Scientifically valid comparison
still requires matching geometry, species content, physical approximations,
resolution, convergence, and definitions of the reported observables.

The \texttt{Diagnostics} branch stores preprocessed arrays used frequently in
analysis and visualization. It is currently populated for GENE records and
can include field and moment data, computational grids, and eigenvalue arrays
such as $k_y$, $\gamma$, and $\omega$. The optional \texttt{Plots} branch
stores convenience visualizations. Neither branch supersedes the native files
or standardized physics representation. The principal standardized data
groups are listed in Supplementary Table~\ref{tab:supp:imas_groups}. Together, the document and binary-file layers preserve model-specific detail,
while the standardized branch exposes the common physics required for
population-scale queries. Next, section~\ref{sec:interfaces} describes how
users interact with these records through command-line, Python, and graphical interfaces.

\section{Access, contribution, and user interfaces}
\label{sec:interfaces}

The value of a shared simulation archive depends not only on how records are
stored, but also on whether they can be contributed, discovered, and retrieved
without bypassing their provenance or standardized representation. MGKDB
therefore exposes the same underlying record model through complementary
interfaces: command-line utilities for repeatable ingestion and retrieval,
Python and PyMongo for programmatic analysis, and MongoDB Compass for
interactive inspection. These interfaces serve different stages of the data
lifecycle rather than representing independent copies of the archive.

The MGKDB software, schema, and client tools are openly available, whereas
access to the curated production database is controlled. Users of the hosted
instance require both access to the NERSC computing environment and an MGKDB
database account. This distinction is important for interpreting the
accessibility of the resource: the software can be deployed locally, while
the production scientific records are shared under managed access conditions.
The repository and documentation provide the current environment,
authentication, and connection procedures.\footnote{MGKDB source and
documentation: \url{https://github.com/Sophelio/MGKDB}.} Operational commands
and connection templates are reproduced in Supplementary
Sec.~\ref{sec:supp:interfaces}.

\begin{table}[tbh]
\centering
\small
\caption{Roles of the principal MGKDB access paths. Capabilities depend on the
permissions associated with the database account. Production writes should be
performed through the validated ingestion and update workflows.}
\label{tab:interface}
\begin{tabularx}{\linewidth}{@{}
    >{\raggedright\arraybackslash}p{0.21\linewidth}
    >{\raggedright\arraybackslash}p{0.31\linewidth}
    >{\raggedright\arraybackslash}X@{}}
\toprule
\textbf{Access path} & \textbf{Primary role} & \textbf{Typical use} \\
\midrule
Command-line utilities & Managed contribution, update, and retrieval & Repeatable batch ingestion; retrieval by collection name or record identifier; scripted archive maintenance \\
Python and PyMongo & Programmatic query and scientific workflow integration & Cohort construction, field projection, diagnostic analysis, visualization, and downstream modeling \\
MongoDB Compass & Interactive discovery and document inspection & Browsing collections, expanding nested fields, testing filters, and inspecting representative records \\
Local MongoDB staging & Preparation of remotely generated or large campaigns & Local validation and organization before a documented merge into the managed production instance \\
\bottomrule
\end{tabularx}
\end{table}

\subsection{Contribution and consistency-preserving updates}
\label{sec:shell}

The command-line uploader is the primary contribution path for supported GENE,
CGYRO, and TGLF records. It accepts code-native directory structures, associates
related files with individual runs, records user-supplied metadata, and invokes
the code-aware translation and diagnostic steps described in
Secs.~\ref{sec:pyrokinetics_context} and~\ref{sec:schema}. The native artifacts
are retained in parallel with the structured document, so ingestion produces a
linked archival record rather than merely extracting a reduced set of arrays.

The source directory organization remains code-aware because the supported
codes emit different file layouts. GENE campaigns commonly distinguish runs by
filename suffixes within a directory, whereas CGYRO and TGLF campaigns commonly
place individual runs in separate subdirectories. The uploader translates
these layouts into the common run-level organization of MGKDB. Current layout
requirements, optional large-file handling, and uploader arguments are given
in Supplementary Sec.~\ref{sec:supp:upload} and maintained in the repository
documentation.\footnote{Upload-format documentation:
\url{https://github.com/Sophelio/MGKDB/wiki/Data-format-for-upload}.}

Records can be amended without discarding their database identity. Metadata-only
changes, such as publication links or clarifying comments, can be applied
separately from changes to the scientific payload. When run-specific native
files are updated through the supported workflow, the standardized
gyrokinetics and applicable diagnostic branches are regenerated so that cached
or translated products do not silently remain inconsistent with the retained
source artifacts. Related calculations can also be linked through database
identifiers, allowing, for example, a reduced-model evaluation to be associated
with a corresponding higher-fidelity run. As emphasized in
Sec.~\ref{sec:schema}, such links establish provenance and pairing; they do not
by themselves establish physical equivalence.

The uploader-assigned review-status field described in
Sec.~\ref{sec:schema} can be supplied during contribution and revised as
additional checks are completed. Because this field is not a substitute for
machine-actionable verification evidence, the comments and provenance fields
should identify the quantities inspected, the convergence procedure, and any
relationship to a parent or restart calculation.

\subsection{Query, retrieval, and programmatic reuse}
\label{sec:python_interface}

MGKDB supports queries at three complementary levels. Metadata queries select
records by code, run class, provenance, workflow, scenario, or review status.
Queries against the IMAS-aligned branch select records by physical content,
such as geometry, species parameters, gradients, growth rates, or transport
outputs. File-level retrieval then recovers the native artifacts associated
with the selected records. Separating discovery from transfer is particularly
useful for nonlinear records: users can first identify a scientifically
coherent cohort from compact structured fields and retrieve large GridFS
payloads only for the cases that require them.

The command-line download utility supports retrieval by folder-derived
\texttt{run\_collection\_name}, MongoDB ObjectID, or individual file
identifier. Because \texttt{run\_collection\_name} is an organizational field
rather than a globally persistent campaign identifier, reproducible analyses
should retain the resolved record identifiers together with the query and
snapshot date. This practice ensures that a frozen cohort remains recoverable
even if the live archive subsequently grows or its human-readable grouping is
revised.

For custom analysis, the MGKDB Python modules and the PyMongo client can be
used from scripts, notebooks, or larger workflow systems.\footnote{PyMongo
documentation: \url{https://pymongo.readthedocs.io/en/stable/}.} Programmatic
access allows users to construct compound filters, request only the fields
needed for an analysis, transform standardized quantities into tabular or
array-based datasets, and retain the originating ObjectIDs as row-level
provenance. It also allows database operations to be embedded in automated
workflows such as SMARTS and PORTALS. The use cases in
Sec.~\ref{sec:case_study} follow this pattern: a database query defines a
cohort, standardized or native content is retrieved, and each derived result
remains traceable to its source record.

Programmatic accessibility should not be confused with unrestricted
comparability. A common query can identify records that share selected fields,
but the resulting cohort must still be screened for compatible models,
normalizations, geometry, species content, resolution, convergence, and
missingness. The interfaces make those checks scalable and auditable; they do
not eliminate them.

\subsection{Interactive inspection and local staging}

\begin{figure}[tbh]
    \centering
    \includegraphics[trim=0cm 3.8cm 0cm 3.8cm, clip,width=0.99\linewidth]{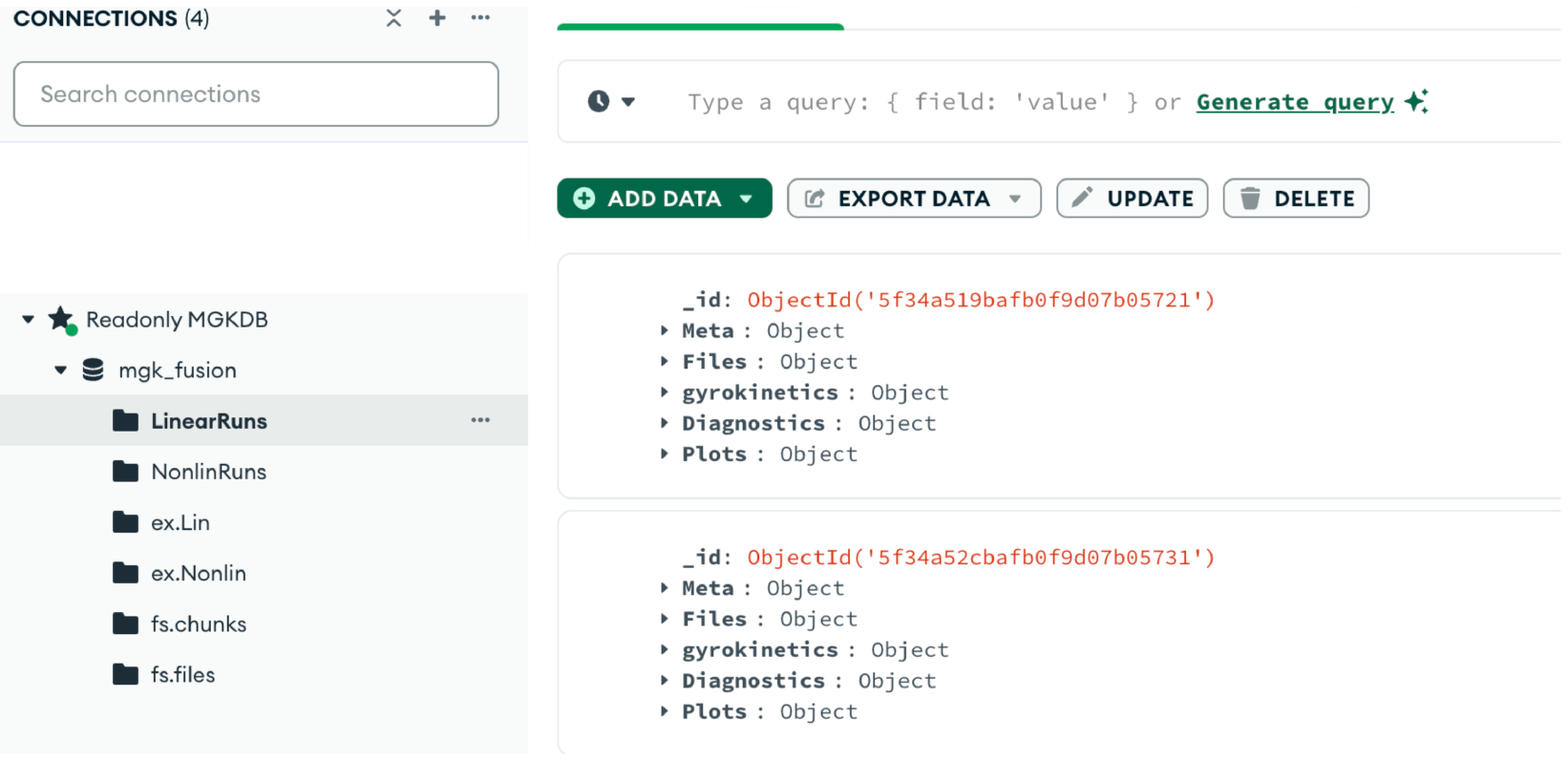}
    \caption{Interactive inspection of MGKDB through MongoDB Compass. The
    collection browser exposes the archive partitions, the filter bar accepts
    MongoDB query expressions, and individual records can be expanded to
    inspect nested metadata, standardized physics fields, and file references.
    The graphical interface is intended primarily for discovery and query
    development; reproducible cohort extraction is performed through saved
    command-line or Python queries.}
    \label{fig:mg_compass_1}
\end{figure}

MongoDB Compass provides a graphical view of the same collections and nested
documents accessed by the command-line and Python interfaces. As illustrated
in Fig.~\ref{fig:mg_compass_1}, users can inspect the collection structure,
expand metadata and scientific branches, and formulate filters interactively.
This is particularly useful when developing a query or determining whether a
field is consistently populated before implementing a larger extraction in
Python. Compass is not the preferred route for bulk file transfer or managed
production updates, and any available write operations remain governed by the
permissions of the connected account.

For campaigns generated on systems from which direct production ingestion is
impractical, the software can also be deployed against a local MongoDB
instance. Records can then be organized and checked locally before transfer
through the documented merge procedure.\footnote{Local-database merge guide:
\url{https://github.com/Sophelio/MGKDB/blob/develop/docs_info/merge_localDB_to_NERSC.md}.}
Local staging therefore extends the ingestion workflow to remote computing
environments without changing the scientific record structure. The merge
itself is an archive-administration operation and should preserve identifiers,
file links, and translated branches according to the current production
procedure.

Taken together, these interfaces connect the architecture of
Secs.~\ref{sec:data_schema} and~\ref{sec:schema} to practical scientific use:
they allow heterogeneous campaign artifacts to enter the archive through a
controlled translation path, allow physically defined populations to be
selected through common fields, and allow the corresponding native or derived
content to be retrieved with record-level provenance. Section~\ref{sec:case_study}
next evaluates these capabilities through database-scale characterization and
cross-fidelity dataset construction.

\section{Database-enabled analyses and reusable modeling workflows}
\label{sec:case_study}

The preceding sections describe how MGKDB preserves code-native artifacts,
translates supported quantities into an IMAS-aligned representation, and
exposes both layers through reproducible query and retrieval interfaces. We
now test what this infrastructure enables scientifically. Here we describe three
demonstrations, which exercise distinct stages of the lifecycle in
Fig.~\ref{fig:generic}: population-scale characterization of standardized
outputs, assessment of archive coverage in a standardized input space, and
construction of a downstream cross-fidelity dataset from retrieved native
files.  An earlier demonstration of database capabilites, using a preliminary MGKDB prototype, was described in~\cite{hatch_22}, which developed reduced models for ETG transport based on archived simulations in MGKDB. 

Each demonstration uses a fixed cohort rather than the continuously changing
production database. The fingerprint analysis uses a July~2026 extract, the
coverage analysis uses an April~26, 2026 extract, and the surrogate workflow
uses a 50-case campaign extract (Table~\ref{tab:snapshot}). The reported
results therefore describe those frozen cohorts. Their purpose is to
demonstrate database-enabled scientific operations, not to establish a final
taxonomy of gyrokinetic modes, an optimal global sampling distribution, or a
production-quality transport surrogate.

\subsection{Population-scale characterization of archived linear modes}
\label{sec:contents}

The first analysis demonstrates how standardized quantities stored in the
Diagnostics branch transform a large collection of linear calculations into
a consistently queryable population for archive-scale physical analysis.  We examine
the archive from two complementary perspectives. The fingerprint analysis
uses mode-resolved outputs to characterize the organization of archived
microinstabilities, while Sec.~\ref{sec:coverage} uses standardized inputs to
measure how the linear archive occupies a common control-parameter space.
Together, they test whether MGKDB supports scientific inference at the level
of an archive rather than only retrieval of individual runs.

\subsubsection{Fingerprint construction and cohort}
\label{sec:fingerprint_quant}

Following Kotschenreuther \textit{et~al.}~\cite{Kotschenreuther2019}, we use several physical `fingerprints’ to classify linear eigenmodes, such as quasilinear transport channels and field-weighted wavenumbers.
The feature family includes growth rate \(\gamma\), real
frequency \(\omega\), electromagnetic-to-electrostatic heat-flux ratios
\(Q_{\mathrm{EM}}/|Q_{\mathrm{ES}}|\) by species, particle-to-heat-flux
ratios \(D/\chi\), parallel parities of \(\widetilde{\phi}\) and
\(\widetilde{A}_{\parallel}\), field-weighted wavenumber moments such as
\(\langle k_{\perp}^{2}\rangle\) and \(\langle k_z\rangle\),
electromagnetic quasilinear weights, and characteristic
field-line widths. These quantities characterize aspects of the underlying instability drive mechanisms as well as the impact on transport channels without depending on the arbitrary amplitude of a linear eigenmode.

The fingerprints are constructed from standardized linear GENE outputs in a
July~2026 database extract and supplemented with a small subset of physical input parameters, including collisionality, and driving gradients. After restricting
the cohort to unstable modes, removing highly collinear columns by
Pearson-correlation screening at \(|r|>0.9\), and applying iterative
variance-inflation-factor screening, the analysis contains
\(N=6{,}053\) modes described by \(d=38\) retained features. Each mode also
carries a preliminary identifier produced by an existing rule-based
classifier: \texttt{ITG\_TEM} (\(2{,}444\)), \texttt{MTM} (\(1{,}126\)),
\texttt{KBM} (\(860\)), \texttt{ETG} (\(659\)), or \texttt{Unknown}
(\(964\)). The \texttt{Unknown} category denotes modes that satisfy none of
the implemented rules and should not be interpreted as a single physical
class.

The rule-derived identifiers and the fingerprint coordinates are not
statistically independent: the rules themselves use quantities from the same
physical descriptor family. The analyses below therefore measure the
geometric organization and reproducibility of this operational taxonomy. They
do not constitute independent physical validation of the assigned mode
identities.

\subsubsection{Embedding and neighborhood agreement}
\label{sec:fingerprint_umap}

\begin{figure}[tbh]
\centering
\includegraphics[width=0.75\linewidth]{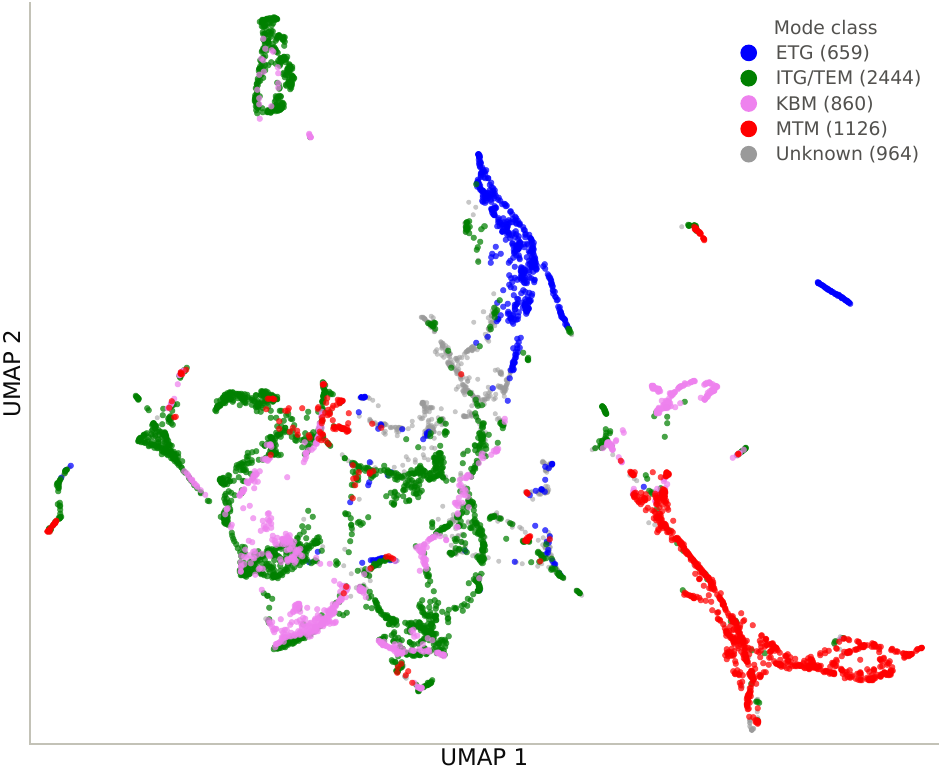}
\caption{UMAP embedding of the transformed and robustly scaled fingerprint
vectors for \(N=6{,}053\) unstable linear GENE modes
(\(n_{\mathrm{nbr}}=30\), \(d_{\min}=0.1\)). Points are colored by their
preliminary rule-derived identifiers, with class sizes reported in the legend;
\texttt{Unknown} modes are shown in light gray. The color assignment follows
the plotting convention used by the FUSE (Fusion Synthesis Engine)
framework~\cite{meneghini2024fuse}; in particular, microtearing modes (MTMs)
are shown in red and kinetic-ballooning modes (KBMs) in magenta. The embedding
is constructed without using these labels and visualizes neighborhood
organization rather than independently validated physical classes.}
\label{fig:fingerprints_umap}
\end{figure}

To visualize the organization of the fingerprint space, features with
absolute sample skewness greater than 2 are transformed with the symmetric-log
mapping used by the analysis code, and all retained features are centered by
their median and scaled by their interquartile range. We then apply UMAP
\cite{McInnes2018UMAP} over
\[
n_{\mathrm{nbr}}\in\{10,20,30,50,80\},
\qquad
d_{\min}\in\{0.05,0.1,0.3\},
\]
where \(n_{\mathrm{nbr}}\) and \(d_{\min}\) denote the UMAP
\texttt{n\_neighbors} and \texttt{min\_dist} parameters. PaCMAP
\cite{Wang2021PaCMAP}, applied to the same transformed features, provides a
projection-method sensitivity check. The representative UMAP view in
Fig.~\ref{fig:fingerprints_umap} uses \(n_{\mathrm{nbr}}=30\) and
\(d_{\min}=0.1\). Class labels are used only after embedding for coloring and
diagnostics; they are not supplied to either manifold-learning algorithm.

The representative embedding contains three prominent regions. ETG modes
form a compact island separated from the other identifiers. MTMs form a second,
elongated region with visible internal structure. ITG/TEM and KBM modes occupy
a broader, partially overlapping ion-scale region into which many
\texttt{Unknown} modes are embedded. PaCMAP and the UMAP parameter sweep
preserve this qualitative organization, although the precise two-dimensional
geometry is projection dependent.

We quantify local label agreement using the fraction of the \(k=30\) nearest
\emph{other} points whose rule-derived label matches that of the focal point.
The corrected same-label statistic preserves the qualitative ordering visible
in the embedding: ETG and MTM have the highest agreement, ITG/TEM is
intermediate, and \texttt{Unknown} and KBM are the most mixed. This result is consistent with~\cite{hatch26}, which identifies MTM as a distinct class from KBM/ITG.  The reduction
relative to a dominant-neighborhood-label statistic is largest for
\texttt{Unknown} and KBM, confirming that a locally homogeneous neighborhood
should not be counted as agreement when it is dominated by a class different
from the focal point. The statistic and the explicit exclusion of the focal
point are defined in Supplementary Sec.~\ref{sec:supp:purity}. No universal
agreement threshold is imposed.

\subsubsection{Projection-independent structure}
\label{sec:fingerprint_cluster}

\begin{figure}[tbh]
\centering
\includegraphics[width=0.55\linewidth]{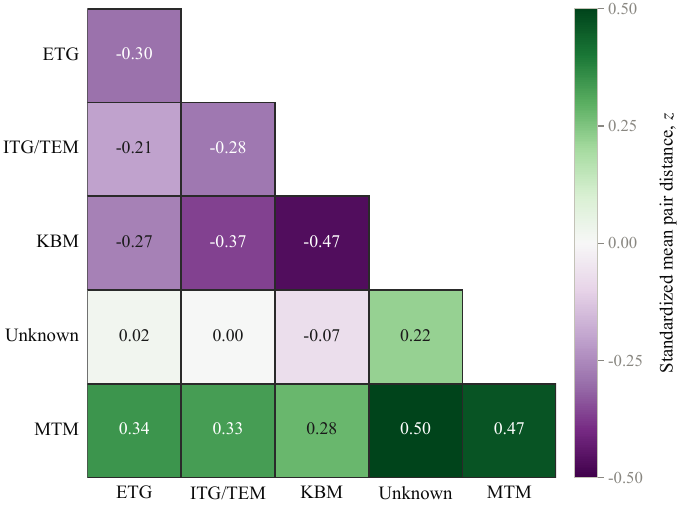}
\caption{Projection-independent comparison of the rule-derived mode groups in
the transformed and scaled fingerprint space. Each cell reports the mean
Euclidean separation for a class pair, standardized against the
database-wide pair-distance distribution. Negative values denote
closer-than-average separation and positive values farther-than-average
separation, in units of the global standard deviation. Outlined diagonal
cells report within-class distances with self-pairs excluded. Only the lower
triangle is shown because the matrix is symmetric.}
\label{fig:standard_scalar_dist}
\end{figure}

Because nonlinear embeddings can distort global distances, we also compare
the classes directly in the transformed 38-dimensional fingerprint space. For
each class pair, the mean Euclidean separation is standardized against the
database-wide pair-distance distribution. Intra-class entries exclude
self-pairs. As shown in Fig.~\ref{fig:standard_scalar_dist}, MTM is separated
from the other rule-derived groups but is internally heterogeneous: every
entry in the MTM row is positive, including an intra-class value of
\(z\simeq+0.47\). A three-component Gaussian-mixture description partitions
the \(1{,}126\) MTMs into groups of \(364\), \(407\), and \(355\) modes with
bootstrap-stability adjusted Rand index \(0.83\). The groups differ primarily
in electromagnetic-to-electrostatic flux ratios, real frequency, and mode
width. They are candidates for further physical investigation, perhaps corresponding to established MTM subtypes, e.g., collisional/collisionless~\cite{predebon_13} or slab/toroidal~\cite{hassan_22}.

ETG, ITG/TEM, and KBM are each internally closer than a typical archive-wide
pair, with diagonal values \(z\simeq-0.30\), \(-0.28\), and \(-0.47\),
respectively. Their pairwise separations are also below the archive-wide
reference, consistent with the proximity of the ion-scale groups in the
embedding. The \texttt{Unknown} population is internally heterogeneous
(\(z\simeq+0.22\)) and is farthest from MTM
(\(z\simeq+0.50\)), reinforcing its interpretation as a residual category
rather than a coherent additional mode family.

Two additional diagnostics test whether the fingerprint representation is
internally recoverable and effectively low dimensional. A cross-validated
one-vs-rest random forest~\cite{breiman2001randomforests} recovers the
rule-derived identifiers with ROC--AUC values of \(0.999\) for ETG, \(0.997\)
for MTM, \(0.991\) for KBM, \(0.986\) for \texttt{Unknown}, and \(0.978\) for
ITG/TEM; the principal five-class confusion is between ITG/TEM and KBM. This
high recovery is an internal consistency result because the labels and
predictors share their physical construction. Separately, the TwoNN
estimator~\cite{facco2017twonn} gives an intrinsic dimension of approximately
\(3.5\), while four and five principal components explain \(90\%\) and
\(95\%\) of the variance, respectively. The 38 retained descriptors therefore
organize along a small number of effective directions within this cohort.

\subsection{Coverage of the standardized input space}
\label{sec:coverage}

The fingerprint analysis characterizes the physics content of stored
eigenmodes; we now consider the complementary question of how the archived
simulations populate their \emph{input} space. Because the extraction uses
quantities represented under common paths in the standardized IMAS branch,
the same field-level query can be applied across supported codes. At the time
of the extraction, the archive contained linear entries from GENE and CGYRO.

From the \textbf{LinearRuns} collection of the April~26, 2026 database
snapshot, we extract every record containing a common set of \(d=13\)
dimensionless control parameters:
\(\beta\), \(q\), \(\hat{s}\), \(\alpha_{\mathrm{MHD}}\),
\(T_i/T_e\), \(k_y\rho_s\), \(R/L_{T_i}\), \(R/L_{T_e}\),
\(R/L_{n_i}\), \(R/L_{n_e}\), \(\nu_{ei}\), and the impurity density
and charge \((n_z/n_e,Z)\). Of the \(26{,}953\) records returned by the
initial query, \(2{,}390\) electron-only simulations are excluded because
\(T_i/T_e\) is undefined, leaving a curated cohort of \(N=24{,}563\)
entries: \(14{,}486\) GENE and \(10{,}077\) CGYRO calculations. Before
computing distances, each parameter is mapped to \([0,1]\) using its empirical
minimum and maximum in the frozen cohort. This scaling prevents parameters
with larger numerical ranges from dominating the Euclidean metric; it does
not imply that the parameters are statistically independent or uniformly
distributed. The normalization and distance definitions are given in
Supplementary Secs.~\ref{sec:supp:coverage_normalization}
and~\ref{sec:supp:coverage_statistics}.

To distinguish the geometry of the archived population from finite-sample
spacing alone, we compare MGKDB with an equally sized Halton low-discrepancy
design~\cite{halton1960} in the same 13-dimensional unit hypercube. The Halton
sample provides an approximately space-filling geometric reference at fixed
\(N\) and \(d\). It is not a physical prior, an expected distribution of
fusion-relevant parameters, or a prescription that the archive should
populate every part of the bounding hypercube uniformly.

\begin{figure}[tbh]
\centering
\includegraphics[width=0.98\linewidth]{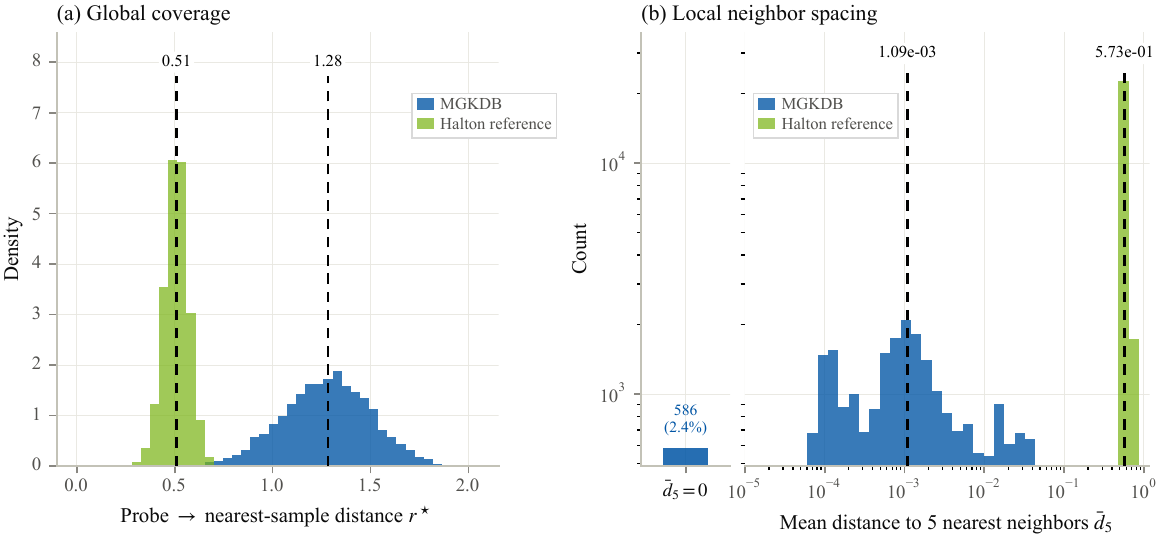}
\caption{Global and local parameter-space coverage of MGKDB
(blue; \(N=24{,}563\) entries represented by \(d=13\) dimensionless
parameters) compared with an equally sized Halton low-discrepancy reference
(green). (a) Distribution of \(r^\star\), the distance from each of
\(M=5{,}000\) uniformly sampled probe points in the normalized
13-dimensional empirical bounding hypercube to the nearest MGKDB or Halton
entry. (b) Distribution of \(\bar{d}_5\), the mean distance from each entry
to its five nearest other entries, restricted to \(\bar{d}_5>0\) for the
logarithmic display. The \(586\) MGKDB entries (\(2.4\%\)) with
\(\bar{d}_5=0\) are shown separately. Dashed vertical lines mark the
distribution medians; in panel~(b), the medians are computed over positive
distances. The Halton design is a fixed-size geometric reference and is not
intended as a physical sampling prior.}
\label{fig:coverage}
\end{figure}

Figure~\ref{fig:coverage} reports complementary global and local diagnostics.
Panel~(a) measures global coverage through
\(r^\star(\mathbf{q})\), the Euclidean distance from each of
\(M=5{,}000\) uniformly sampled probe points \(\mathbf{q}\) to its nearest
entry. The MGKDB distribution is broad, with median
\(r^\star\simeq1.28\), approximately \(2.5\) times the Halton-reference
median of \(r^\star\simeq0.51\). Thus, a typical point in the empirical
13-dimensional bounding hypercube lies substantially farther from an archived
simulation than it would under an equally sized space-filling design.

The principal-component spectrum provides a complementary explanation for
this result. Four principal components capture \(80\%\) of the variance and
five capture \(87\%\), while the participation ratio of the spectrum gives
approximately \(4.5\) effective dimensions. The archived population
therefore occupies a structured, lower-dimensional subset of its
13-dimensional bounding box rather than filling that box uniformly. This
compression reflects both relationships among the physical inputs and the
campaigns from which the records were contributed. For example, the gradient
drives are strongly correlated:
\(R/L_{n_e}\) and \(R/L_{n_i}\) have \(r=0.98\), while
\(R/L_{T_e}\) and \(R/L_{n_e}\) have \(r=0.97\). These correlations can arise
from the profile construction, physical constraints, and scan strategies used
in the originating campaigns and should not be interpreted as universal
relationships over all physically admissible plasmas.

Panel~(b) examines local concentration using \(\bar{d}_5\), the mean distance
from each entry to its five nearest other entries. Positive distances are
displayed on logarithmic axes. The Halton reference is concentrated near
\(\bar{d}_5\simeq0.57\), whereas the positive-distance MGKDB distribution is
shifted by roughly three orders of magnitude toward smaller separations, with
median
\[
\bar{d}_5\simeq1.09\times10^{-3}.
\]
In addition, \(586\) MGKDB entries (\(2.4\%\)) have numerically zero
\(\bar{d}_5\) and are displayed separately from the logarithmic histogram.
For such an entry, all five nearest neighbors occupy the same point in the
retained 13-parameter representation. These \(586\) entries therefore belong
to configurations containing at least six coincident entries and form a
stricter subset of the population containing repeated parameter vectors.

The local concentration has a concrete connection to campaign and scan
structure. When the scanned wavenumber \(k_y\rho_s\) is omitted, the
\(24{,}563\) entries collapse onto only \(4{,}101\) distinct configurations,
or approximately six \(k_y\) values per configuration. Even when
\(k_y\rho_s\) is retained, \(1{,}781\) rows repeat a 13-parameter vector that
appeared earlier in the extracted table; equivalently, the \(24{,}563\) rows
contain \(22{,}782\) distinct 13-parameter vectors. The counting conventions
and their relationship to the \(586\) zero-distance entries are detailed in
Supplementary Sec.~\ref{sec:supp:coverage_duplicates}. This statement
concerns duplication in the retained parameter representation, not
necessarily duplication of complete database records. A gyrokinetic
calculation contains additional information---including higher-order
geometry, species composition, collision models, numerical resolution,
provenance, and workflow context---that is not captured by these 13 scalars.
Consequently, two scientifically distinct calculations can project onto the
same point or onto nearby points in the reduced control space.

These global and local structures should therefore be interpreted as
properties of the archive rather than automatically as data defects. The
global concentration records the physical relationships and campaign choices
that shaped the database, while the local concentration reflects deliberate
parameter scans, repeated operating points, and the finite resolving power of
the selected descriptor. Nevertheless, coincident parameter vectors are not
statistically independent in this representation. Machine-learning and
statistical applications must account for campaign membership and repeated
configurations when defining training and validation partitions; otherwise,
closely related or identical parameter vectors can be split across folds,
inflate apparent generalization, or overweight heavily represented campaigns.

A complementary campaign-planning question is which coordinates merit greater
sampling in future contributions. We construct a heuristic priority score from
the product of two factors: the contribution of each parameter to the leading
principal components, measured by its squared loadings weighted by explained
variance, and its marginal under-sampling, measured by one minus the normalized
entropy of its empirical histogram. A parameter scores highly only when it
both contributes substantially to the populated data manifold and is sampled
unevenly within the current archive. The definition and limitations of this
score are given in Supplementary
Sec.~\ref{sec:supp:coverage_ranking}.

Under this diagnostic, impurity charge \(Z\) is the highest-ranked coordinate,
followed by the four gradient drives
\(R/L_{n_e}\), \(R/L_{n_i}\), \(R/L_{T_e}\), and \(R/L_{T_i}\).
The impurity-charge result requires special care: the archived values are
effectively categorical, with \(Z\in\{0,4,6\}\) and carbon accounting for
\(94\%\) of the entries. Its high score therefore indicates limited impurity
variety rather than a continuous interval that should simply be filled more
densely. Among the continuous parameters, the ranking identifies the gradient
drives as influential but unevenly sampled directions in the current archive.

This score is not an optimal experimental-design criterion and does not, by
itself, determine which simulation should be run next. Its value is as an
archive-level screening tool: combined with physical objectives, model
validity, computational cost, and campaign provenance, it identifies
directions in which new simulations are most likely to broaden the information
content of MGKDB. Recomputing the diagnostic as the archive grows would provide
an evolving, quantitative basis for coordinating future contributions.

\subsection{Implications for similarity search and campaign design}
\label{sec:implications}

The output- and input-space analyses suggest complementary operational uses
for the archive. In fingerprint space, the nearest neighbors of a newly
ingested linear GENE mode provide candidate reference cases for inspecting
mode structure, transport channels, and parameter sensitivity. Because each
point remains linked to its complete MGKDB record, an embedding can serve as a
navigation or similarity layer over the archive. Neighbor proximity alone is
not evidence of code validation or physical equivalence; candidate comparisons
must still be screened using the model and provenance information described in
Sec.~\ref{sec:schema}.

In input space, \(r^\star(\mathbf{q})\) supplies a low-cost indication that a
proposed operating point lies far from the populated MGKDB cohort under the
selected 13-parameter representation. Such a point may merit a new simulation,
but a large distance can also reflect a physically inadmissible combination or
a direction omitted from the descriptor. Conversely, a small distance does
not imply redundant physics, because records close in the 13-dimensional space
can differ in geometry, species treatment, resolution, or other retained
metadata. Coverage diagnostics should therefore be combined with domain
constraints and record-level provenance when selecting future calculations.

The repeated-vector census also has direct implications for data-driven work.
Random row-level splits can place identical parameter vectors or related
members of the same campaign in both training and test sets. Depending on the
prediction target, this can inflate apparent generalization and cause large
campaigns to dominate model fitting. Dataset builders should use grouped
splits, duplicate-aware weighting, or explicit campaign holdouts appropriate
to the scientific question. MGKDB makes these controls possible by keeping
the standardized coordinates connected to record identifiers, campaign
groupings, and native artifacts.

\subsection{From archive to a working surrogate: a CGYRO--TGLF proof of concept}
\label{sec:surrogates}

The final demonstration treats MGKDB as an active component of a modeling
workflow. It has two stages: an automated pipeline that constructs TGLF
evaluations at operating points recovered from archived nonlinear CGYRO runs,
and a small surrogate trained on the resulting reduced-model targets. The aim
is to demonstrate reproducible dataset construction across model fidelities,
not to benchmark surrogate architectures or validate TGLF against CGYRO.

\subsubsection{Constructing reduced-model evaluations from archived runs}
\label{sec:cross_fidelity}

The archived nonlinear CGYRO records preserve the input specifications of
calculations that required substantial computational effort. Using
\textit{Pyrokinetics}~\cite{patel2024pyrokinetics}, the workflow converts a
retrieved \texttt{input.cgyro} file into a corresponding TGLF input and
executes the reduced quasilinear model~\cite{tglf}. It then extracts TGLF
growth-rate spectra \(\gamma(k_y)\), real frequencies, and quasilinear flux
estimates. The result is a set of reduced-model outputs evaluated at operating
points inherited from the archived high-fidelity campaign. These are matched
operating points, not automatically matched observables: direct physical
assessment of TGLF would require a separately defined comparison with
commensurate nonlinear CGYRO quantities.

Candidate records are selected through a metadata query, including
\texttt{sim\_type = CGYRO} and \texttt{IsLinear = False}, and their native
inputs are retrieved with the MGKDB client distributed as
\texttt{MGKDB-fusion}. The demonstration uses 50 nonlinear CGYRO operating
points from one ITER-startup PORTALS optimization campaign
\cite{rodriguezfernandez2024portals}: ten optimizer iterations at each of five radial
locations, \(\rho\in\{0.35,0.55,0.75,0.88,0.90\}\). Query, retrieval,
conversion, TGLF execution, and feature extraction are automated end to end.

\subsubsection{Proof-of-concept surrogate construction}
\label{sec:surrogate_poc}

For each operating point, the workflow constructs a 38-component input vector
spanning Miller-geometry coefficients, per-species density and temperature
gradients, \(\beta_e\), collisionality, \(Z_{\mathrm{eff}}\), and rotation
quantities. These features are paired with a hierarchy of TGLF targets. The
primary estimator is ridge regression~\cite{hoerl1970ridge}, with feature
standardization and model fitting performed independently inside every
training fold. A small multilayer perceptron provides a nonlinear comparison.
Further model and scoring details are given in Supplementary
Secs.~\ref{sec:supp:surrogate_model} and~\ref{sec:supp:surrogate_scoring}.

\begin{figure}[tbh]
\centering
\includegraphics[width=0.95\linewidth]{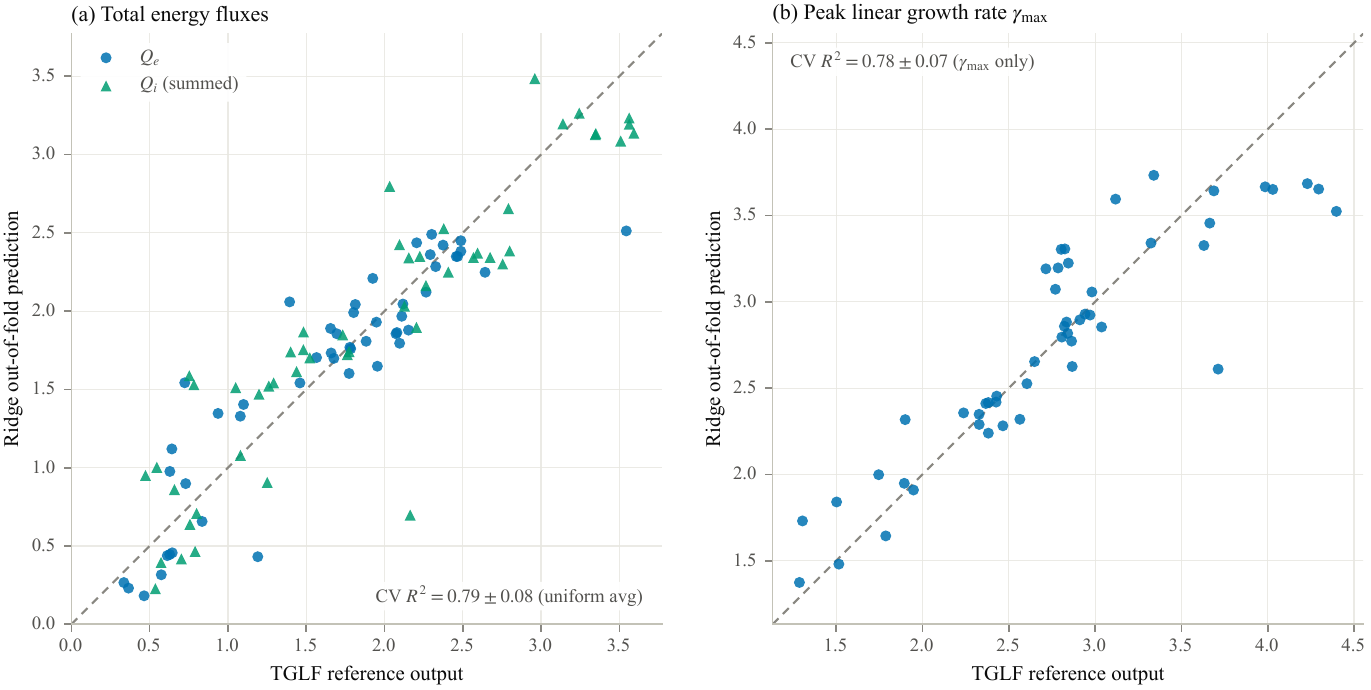}
\caption{Out-of-fold parity plots for the ridge surrogate under shuffled
five-fold cross-validation on 50 archived CGYRO operating points. (a) Electron
and summed-ion energy fluxes in gyro-Bohm units. (b) Peak linear growth rate
\(\gamma_{\max}\), corresponding to component~0 of the three-component
instability-peak target. Horizontal axes show TGLF reference outputs and
vertical axes show ridge-model predictions; dashed lines denote equality. The
panel~(a) annotation reports the uniform-average \(R^2\) across its two
displayed flux outputs, whereas panel~(b) reports the component-specific
\(R^2\) for \(\gamma_{\max}\). Values are the mean and population standard
deviation across folds. The scores quantify within-dataset learnability and
do not estimate transfer to unseen campaigns or radial locations.}
\label{fig:surrogate_parity}
\end{figure}

All displayed points are out-of-fold predictions from shuffled five-fold
cross-validation. Under this within-dataset evaluation, lower-dimensional
targets are learned more successfully than higher-dimensional ones. The
uniform-average score across the electron and summed-ion energy fluxes is
\(R^2=0.79\pm0.09\) (Fig.~\ref{fig:surrogate_parity}(a)), while the
component-specific score for the peak growth rate is
\(R^2=0.78\pm0.07\) (Fig.~\ref{fig:surrogate_parity}(b)). Across the broader
target hierarchy, the per-species energy fluxes attain
\(R^2=0.79\pm0.08\), the complete three-component instability-peak target
\((\gamma_{\max},k_y^{\max},\omega_{\max})\) attains
\(R^2=0.64\pm0.07\), and the 33-component \(\gamma(k_y)\) spectrum attains
\(R^2=0.56\pm0.08\). Values are means and population standard deviations of
the five foldwise scores; multi-output scores use a uniform average across
target components. Raw high-dimensional targets, including the 132-component
eigenvalue spectrum, yield negative cross-validated \(R^2\) at this sample
size, and the multilayer perceptron underperforms the regularized linear model.

The random folds can place optimizer iterations and radial locations from the
same campaign in both the training and test sets. The scores therefore show
that the automated pipeline produces internally coherent feature--target pairs;
they do not establish generalization beyond this 50-case campaign. More
stringent tests would hold out all radial points from one optimizer iteration,
hold out one radius across iterations, and ultimately hold out complete
campaigns. Likewise, the surrogate emulates TGLF outputs rather than nonlinear
CGYRO transport. Establishing reduced-model fidelity requires direct,
observable-matched TGLF--CGYRO comparisons and is outside the scope of this
demonstration.

The principal result is the transformation of an archived campaign into an
analysis-ready cross-fidelity dataset through a reproducible sequence of
database and model operations. The same pattern can be extended where
\textit{Pyrokinetics} supplies the required readers, writers, and normalization
mappings. MGKDB thereby lowers the data-engineering cost of constructing such
datasets while retaining the provenance needed to determine what a trained
model actually represents.

Taken together, the three demonstrations connect the system objectives of
Sec.~\ref{sec:overview} to observable operations. Standardized outputs support
population-scale characterization of \(6{,}053\) unstable modes;
standardized inputs support coverage analysis of \(24{,}563\) eligible linear
records; and retained native files support automated cross-fidelity dataset
construction from 50 nonlinear operating points. The demonstrations are not
formal benchmarks of database latency, ingestion throughput, mode
classification, experimental design, or surrogate transfer. They show that a
heterogeneous archive can be converted into traceable scientific populations
and reusable workflows without severing derived results from their source
records.

\section{Discussion and outlook}
\label{sec:outlook}

MGKDB addresses a practical obstacle to cumulative fusion modeling: expensive
simulations are often preserved as campaign-specific directory structures
rather than as a scientific population that can be searched, compared, and
reused. The central design choice is to retain two complementary
representations of each supported calculation. Code-native artifacts preserve
model-specific information and provide the strongest available basis for
inspection and reproduction, while the IMAS-aligned branch exposes selected
physics quantities through common paths and normalization conventions. Neither
representation is sufficient by itself. Native files without structured
metadata remain difficult to discover and combine, whereas standardized arrays
without their originating files and provenance can obscure assumptions needed
to interpret them. Keeping both layers connected is therefore the principal
engineering contribution of MGKDB.

\subsection{What the demonstrations establish}
\label{sec:discussion_demonstrations}

The three demonstrations in Sec.~\ref{sec:case_study} test this architecture at
different points in the data lifecycle. The fingerprint analysis shows that
standardized mode-resolved outputs can support population-scale analysis across
thousands of archived calculations. The coverage analysis shows that common
input coordinates can reveal the geometry, campaign structure, and repeated
configurations of a multicode archive. The CGYRO--TGLF workflow shows that
record-level queries can recover native files, drive an external model, and
produce a traceable dataset for downstream learning. Together, these examples
move the contribution beyond file storage: they demonstrate a path from
archived calculations to reproducible scientific operations.

The demonstrations also expose requirements that a useful simulation archive
must satisfy. Population analyses require explicit cohort definitions and
dated snapshots because the production database continues to evolve.
Coverage studies require attention to repeated parameter vectors and campaign
structure because archived runs are not independent samples from a uniform
distribution. Machine-learning workflows require group-aware validation and
observable-matched targets because random row-level partitions can exaggerate
generalization. Cross-code and cross-fidelity studies require model assumptions
and native provenance to remain available because a common field name does not
establish physical equivalence. These are not secondary caveats; they are
design requirements for scientifically defensible reuse.

The results should therefore be interpreted as capability demonstrations rather
than performance benchmarks. The fingerprint analysis evaluates the internal
organization of an operational, rule-derived taxonomy rather than independently
validating mode identities. The coverage analysis characterizes the populated
archive relative to a geometric reference rather than defining an optimal
distribution of fusion simulations. The surrogate reproduces selected TGLF
outputs within one small campaign rather than predicting nonlinear CGYRO
transport or establishing transfer to unseen campaigns. This narrower
interpretation strengthens the paper's principal claim: MGKDB provides the
infrastructure needed to formulate, audit, and progressively strengthen such
analyses.

\subsection{Interoperability without loss of model context}
\label{sec:discussion_interoperability}

The common representation is most valuable as a discovery and alignment layer.
It allows users to ask physically formulated questions across supported records
without first writing a parser for every native format. IMAS supplies the
shared scientific organization~\cite{Imbeaux_2015}, while
\textit{Pyrokinetics} supplies code-aware reading, writing, and normalization
transformations for the currently supported gyrokinetic and reduced-model
workflows~\cite{patel2024pyrokinetics}. The continued development and
open-source availability of these components
\cite{iter_imas_open_2025,fair4rs} make it possible for MGKDB to participate
in a broader ecosystem of integrated fusion modeling rather than defining an
isolated database-specific format.

Standardization nevertheless has a clear boundary. Two records that expose the
same standardized field may still differ in geometry, species content,
collisions, numerical resolution, saturation assumptions, convergence, or
domain of validity. The distinction is especially important across fidelity
levels. TGLF is a reduced quasilinear model, whereas GENE and CGYRO solve
gyrokinetic equations at higher fidelity; evaluating them at the same operating
point does not make their outputs interchangeable. MGKDB can identify candidate
comparisons and preserve the evidence needed to assess them, but verification
and validation remain scientific activities rather than database properties
\cite{oberkampf2004verification}.

This distinction also defines a disciplined route to broader model coverage.
The document architecture can accommodate reduced-fluid and
magnetohydrodynamic calculations, including frameworks such as
BOUT++~\cite{dudson2009bout}, but storage compatibility alone is not sufficient
for production support. Each additional model family requires a defined native
payload, model-specific provenance, validated readers, explicit normalization
and coordinate mappings, appropriate standardized fields, and tests that
establish the limits of cross-model comparison. Expansion should therefore be
driven by scientifically meaningful mappings rather than by the goal of placing
heterogeneous outputs under a single nominal schema.

\subsection{Current scope and limitations}
\label{sec:discussion_limitations}

The current archive has several limitations that should guide its use and
development. First, its population reflects contributed campaigns rather than
a statistically designed survey of fusion-relevant parameter space. The
coverage results show strong correlations, scan structure, and repeated
configurations. These features are scientifically informative, but they mean
that archive-wide statistics can be dominated by a small number of workflows
unless campaign identity and sampling multiplicity are incorporated explicitly.

Second, metadata and retained payloads are not perfectly uniform across the
history of the database. Legacy records can lack fields introduced by later
ingestion pipelines, and the presence of a standardized branch does not
guarantee that every possible quantity was available or translated. Likewise,
the uploader-assigned \texttt{confidence} field records review status; it is
not a probability of correctness or a certificate that all outputs are
converged and physically valid. Analyses must test field availability,
document exclusion criteria, and select validation evidence appropriate to
their observable.

Third, preservation does not guarantee indefinite executability. Historical
native files can remain interpretable while the corresponding code version,
compiler, dependency stack, or execution environment becomes difficult to
reconstruct. Exact reruns may therefore require source snapshots, environment
specifications, containers, or external licenses that are not presently
guaranteed for every record. MGKDB preserves the strongest available run-level
evidence, but it should not be described as providing bitwise reproducibility
for all historical calculations.

Finally, the software and hosted data have different access conditions. The
MGKDB source code, schema, and client interfaces are open, whereas the curated
production archive requires NERSC authorization and MGKDB credentials. This is
managed accessibility rather than unrestricted public release. FAIR practice
requires that conditions of access be explicit and that reusable data and
software retain sufficient metadata; it does not require every research object
to be anonymously downloadable~\cite{fair,fair4rs}. Nevertheless, broader
community use would benefit from well-documented benchmark cohorts that can be released when contributor permissions and data policies allow.

\subsection{Schema evolution, versioning, and reproducibility}
\label{sec:evolution}

Long-lived simulation archives must manage several independent forms of
versioning: the source simulation code and its native format, the database
record schema, the IMAS data-model release, the translation software, and the
analysis code applied to a frozen cohort. Conflating these layers makes it
difficult to determine whether a changed value reflects new physics, a revised
parser, a normalization correction, or a database migration. Future records
should therefore expose these versions separately and retain the conversion
history needed to reconstruct the standardized representation.

The IMAS schema has already changed during the lifetime of MGKDB. A migration
now in progress updates entries created against IMAS version~3.28 to the
release supported by the current ingestion pipeline. Such migrations should
regenerate translated or derived branches while preserving the original
code-native artifacts. At minimum, the record should retain the originating
schema version, target schema version, translation-software version, migration
timestamp, and any validation result. This approach permits common-field
queries to evolve without erasing the representation from which an archived
result was derived.

The same principle applies to scientific analyses. A reproducible published
cohort should be documented by its query or selection definition, extraction
date, analysis-software revision, environment specification, random seeds,
and numerical conventions. Where access policies permit, resolved record
identifiers and relevant file or record checksums provide additional
provenance. This documentation distinguishes dated analyses from the evolving
production archive and aligns the data and software components of MGKDB more
closely with the FAIR principles~\cite{fair,fair4rs}.

\subsection{Integration with external workflows}
\label{sec:integration}

MGKDB is distributed on PyPI under the package name
\texttt{MGKDB-fusion}, with the Python import namespace \texttt{mgkdb}. The
supported client installation is

\begin{quote}
\small\ttfamily
pip install MGKDB-fusion
\end{quote}

\noindent This installation exposes the upload, query, retrieval, and
record-processing interfaces for use in scripts, notebooks, and automated
workflows. The source repository can additionally be cloned when a user needs
the development version, examples, documentation, or repository-level
utilities. Using one canonical package name and repository URL in the paper,
documentation, and package metadata is important because similarly named
historical packages can otherwise lead users to unsupported installations.

The most productive integrations will preserve record identity across the
entire workflow. A transport-optimization system, for example, should retain
the MGKDB identifiers of the simulations used to initialize or train a model,
the conversion version used to construct its inputs, and the identifiers of
new calculations returned to the archive. This pattern complements integrated
modeling environments built around IMAS-compatible data
\cite{Meneghini_2021,lyons2023step} and surrogate-assisted transport workflows
\cite{vandeplassche2020,rodriguezfernandez2024portals}. The database then functions
not merely as an input store, but as a provenance layer connecting simulation,
optimization, reduced modeling, and learned representations.

\subsection{Priorities for community-scale development}
\label{sec:discussion_roadmap}

The immediate development priorities follow directly from the limitations
identified above. First, ingestion should produce increasingly
machine-actionable provenance: stable campaign identifiers, checksums for
native artifacts, explicit code and translation versions, and structured
verification results. Second, schema migrations should be accompanied by
automated field-level validation and migration reports. Third, benchmark
cohorts should define duplicate-aware and campaign-aware splits so that
machine-learning results can be compared without hidden overlap between
training and evaluation data.

Scaling the archive will also require operational engineering. Nonlinear
campaigns can move MGKDB from hundreds of gigabytes toward multi-terabyte data
volumes, increasing the importance of selective retrieval, compact metadata
indexes, storage-policy testing, integrity checks, and measured performance
under representative access patterns. These system-level benchmarks are
distinct from the scientific demonstrations in Sec.~\ref{sec:case_study} and
should be reported separately as the production service grows.

Finally, the archive will benefit from a contribution and governance model in
which new codes, model families, and diagnostics are admitted through
documented mappings and validation tests. Community stewardship is essential:
the value of MGKDB depends not only on record count, but on whether contributors
can understand the schema, identify the provenance of a result, report
translation problems, and cite stable datasets. Periodic public benchmark
releases, where permissions permit, would lower the barrier to independent
method development while the managed production archive continues to support
larger or restricted campaign data.

The broader opportunity is to make simulation campaigns cumulative. A
calculation contributed for one study can later participate in a cross-code
comparison, identify a gap in parameter coverage, seed a reduced-model
evaluation, or support a new data-driven method without losing its original
context. MGKDB does not remove the scientific judgment required for those
uses; it makes that judgment more scalable, explicit, and auditable. That
combination---standardized discovery, preserved native evidence, and
record-level provenance---is what can turn the accumulated computational cost
of fusion modeling into durable community infrastructure.

\section{Conclusions}
\label{sec:conclusions}

MGKDB converts heterogeneous fusion-simulation outputs into traceable
scientific records that can be discovered, inspected, and reused without
discarding their model-specific context. The system combines an open-source
software framework with a curated production archive at NERSC and links
code-native artifacts to provenance and quality metadata, an IMAS-aligned
physics representation, and derived diagnostics. At the September~1, 2026 reference snapshot, the archive contained
\(1{,}068{,}089\) records, including linear and nonlinear GENE and CGYRO
calculations and reduced quasilinear TGLF evaluations; \(1{,}067{,}980\)
records (\(99.99\%\)) contained a populated gyrokinetics IMAS branch. The resulting architecture preserves the information
needed to interpret an individual calculation while allowing compatible
records to participate in common queries and population-scale workflows.

The three demonstrations show how these linked representations support
distinct scientific operations. Standardized mode-resolved outputs enabled
the organization of \(6{,}053\) unstable linear GENE modes to be examined in a
physics-motivated fingerprint space. Common input coordinates allowed
\(24{,}563\) linear GENE and CGYRO records to be assessed for global coverage,
local concentration, repeated configurations, and campaign-driven sampling
structure. Retrieval of native CGYRO inputs enabled TGLF evaluations to be
generated at the same operating points as an archived nonlinear CGYRO
campaign, producing an analysis-ready dataset for exploratory surrogate
modeling. These studies are deliberately
scoped capability demonstrations rather than definitive benchmarks of mode
classification, optimal campaign design, reduced-model fidelity, or surrogate
transfer. Their shared result is that MGKDB can connect archive-scale analysis
and downstream modeling to identifiable source records.

The lasting value of MGKDB will depend less on record count alone than on the
quality of this connection. Versioned translations, explicit provenance,
stable campaign identifiers, well-documented fixed cohorts, validation-aware contribution pathways, and scientifically defined mappings for additional
model families will determine whether the archive remains interpretable as it
grows. Developed in this way, MGKDB can provide a common infrastructure through
which previously isolated simulations support code comparison, coverage
assessment, reduced-model development, machine learning, and the design of new
computational campaigns. Its broader purpose is to make fusion simulation
cumulative: to ensure that the computational investment in one calculation or
campaign remains available as evidence and input for the next.

\setcounter{section}{0}
\setcounter{subsection}{0}
\setcounter{subsubsection}{0}
\setcounter{figure}{0}
\setcounter{table}{0}
\setcounter{equation}{0}

\renewcommand{\thesection}{S\arabic{section}}
\renewcommand{\thesubsection}{\thesection.\arabic{subsection}}
\renewcommand{\thesubsubsection}{\thesubsection.\arabic{subsubsection}}
\renewcommand{\thefigure}{S\arabic{figure}}
\renewcommand{\thetable}{S\arabic{table}}
\renewcommand{\theequation}{S\arabic{equation}}

\renewcommand*{\theHsection}{supp.section.\arabic{section}}
\renewcommand*{\theHsubsection}{supp.section.\arabic{section}.\arabic{subsection}}
\renewcommand*{\theHsubsubsection}{supp.section.\arabic{section}.\arabic{subsection}.\arabic{subsubsection}}
\renewcommand*{\theHfigure}{supp.figure.\arabic{figure}}
\renewcommand*{\theHtable}{supp.table.\arabic{table}}
\renewcommand*{\theHequation}{supp.equation.\arabic{equation}}

\pagebreak

\section*{Supplementary material}
\addcontentsline{toc}{section}{Supplementary material}

This Supplementary Material provides the cohort definitions, mathematical
conventions, methodological details, and implementation-level documentation
supporting the Multiscale GyroKinetic DataBase (MGKDB) architecture and the
three database-enabled demonstrations reported in the main manuscript.
Supplementary Sec.~\ref{sec:supp:snapshots} records the dated database
snapshots and frozen analysis cohorts; Supplementary
Secs.~\ref{sec:supp:fingerprints}--\ref{sec:supp:surrogate} document the
fingerprint-space characterization, parameter-space coverage analysis, and
CGYRO--TGLF proof-of-concept workflow presented in
Sec.~\ref{sec:case_study}; and Supplementary
Secs.~\ref{sec:supp:schema_reference} and~\ref{sec:supp:interfaces} provide
the detailed record-schema and operational-interface material supporting
Secs.~\ref{sec:data_schema} and~\ref{sec:interfaces}, respectively.
Quantities, acronyms, and scoring conventions required to interpret the
supplementary methods are defined at first use; citations are included in the
common bibliography of the article.

\section{Analysis snapshots and cohort provenance}
\label{sec:supp:snapshots}

All numerical results reported in the manuscript are tied to fixed database
snapshots or frozen analysis cohorts rather than to the continuously evolving
production archive. Table~\ref{tab:supp:snapshots} distinguishes the
production snapshot used for archive-wide census statements from the three
cohorts used for the scientific demonstrations. These cohorts were constructed
independently and are not assumed to be subsets of the September~1, 2026
production snapshot. They therefore should not be combined to infer the
composition of MGKDB at a single point in time.

\begin{table}[htb!]
\centering
\small
\caption{Production snapshot and frozen analysis cohorts used in the
manuscript. The retained-sample column identifies the relevant counting unit,
which differs among records, individual linear modes, and operating points.}
\label{tab:supp:snapshots}

\begingroup
\setlength{\tabcolsep}{4pt}
\renewcommand{\arraystretch}{1.16}

\begin{tabularx}{\linewidth}{@{}
    >{\raggedright\arraybackslash}p{0.17\linewidth}
    >{\raggedright\arraybackslash}p{0.19\linewidth}
    >{\raggedright\arraybackslash}X
    >{\raggedright\arraybackslash}p{0.14\linewidth}@{}}
\toprule
\textbf{Purpose} &
\textbf{Snapshot or cohort} &
\textbf{Selection definition} &
\textbf{Retained sample} \\
\midrule

Production census &
September~1, 2026 &
All records present in the \texttt{LinearRuns} and
\texttt{NonlinRuns} collections. &
\(1{,}068{,}089\) records \\

Fingerprint analysis &
July~2026 frozen extract &
Unstable linear GENE modes retained after fingerprint extraction and
quality filtering. &
\(6{,}053\) modes \\

Coverage analysis &
April~26, 2026 snapshot &
Linear GENE and CGYRO records in \texttt{LinearRuns}. Of the
\(26{,}953\) records initially queried, \(2{,}390\) electron-only
calculations were excluded because \(T_i/T_e\) was undefined, leaving
records with all 13 required control parameters. &
\(24{,}563\) records \\

Surrogate demonstration &
PORTALS campaign cohort &
Nonlinear CGYRO operating points from one ITER-startup optimization
campaign, comprising ten optimizer iterations at each of five radial
locations; TGLF was then evaluated at the same operating points. &
\(50\) operating points \\

\bottomrule
\end{tabularx}

\endgroup
\end{table}

Each frozen cohort is defined by the source collection and extraction
timestamp, the complete database filter and field projection, the resolved
MGKDB record identifiers, and all post-query exclusions. Reproducing the
reported results additionally requires the analysis-code commit, software
environment, random seeds, transformation and scoring conventions, and the
generated numerical artifacts underlying the tables and figures. 

The live production archive may change after the reference date without
altering results computed from these frozen cohorts. Conversely, the counts
reported for a dated cohort should not be interpreted as current production
counts or as evidence of temporal changes between independently constructed
extracts.

\section{Fingerprint-space analysis}
\label{sec:supp:fingerprints}

The fingerprint analysis examines whether standardized mode-resolved outputs
expose reproducible geometric organization across a heterogeneous population
of archived linear calculations. Following the physical-descriptor framework
of Kotschenreuther \textit{et al.}~\cite{Kotschenreuther2019}, the analysis
uses dimensionless ratios and field-weighted quantities that characterize mode
structure and quasilinear transport response without depending on the
arbitrary amplitude of a linear eigenmode.

The frozen cohort contains \(N=6{,}053\) unstable linear GENE modes. Each mode
has one preliminary rule-derived identifier:
\texttt{ITG\_TEM} (\(2{,}444\) modes), \texttt{MTM} (\(1{,}126\)),
\texttt{KBM} (\(860\)), \texttt{ETG} (\(659\)), or \texttt{Unknown}
(\(964\)). The \texttt{Unknown} category contains modes that satisfy none of
the implemented classification rules and is not assumed to constitute a
single physical mode family. Because the classification rules use quantities
from the same descriptor family as the fingerprint coordinates, the analyses
below test the organization and reproducibility of this operational taxonomy;
they do not provide independent physical validation of the assigned labels.

\subsection{Feature preparation and embeddings}
\label{sec:supp:fingerprint_preprocessing}

Highly collinear candidate features are removed using Pearson-correlation
screening at \(\lvert r\rvert>0.9\), followed by iterative
variance-inflation-factor screening, leaving \(d=38\) retained features.
Features with absolute sample skewness greater than \(2\) are transformed with
the symmetric-log mapping implemented in the frozen analysis workflow. Every
retained feature is then centered by its sample median and divided by its
interquartile range. The resulting transformed and scaled fingerprint vector
for mode \(i\) is denoted by
\(\mathbf{x}_i\in\mathbb{R}^{38}\).

Uniform Manifold Approximation and Projection
(UMAP)~\cite{McInnes2018UMAP} is evaluated over
\begin{equation}
n_{\mathrm{nbr}}\in\{10,20,30,50,80\},
\qquad
d_{\min}\in\{0.05,0.1,0.3\},
\label{eq:supp:umap_sweep}
\end{equation}
where \(n_{\mathrm{nbr}}\) and \(d_{\min}\) denote the UMAP
\texttt{n\_neighbors} and \texttt{min\_dist} parameters, respectively.
The representative two-dimensional visualization in
Fig.~\ref{fig:fingerprints_umap} uses
\(n_{\mathrm{nbr}}=30\) and \(d_{\min}=0.1\). Euclidean distance is used in
the transformed feature space when constructing the embedding. The
rule-derived labels are not supplied to UMAP and are used only for coloring
and subsequent diagnostics.

Pairwise Controlled Manifold Approximation Projection
(PaCMAP)~\cite{Wang2021PaCMAP} is applied to the same transformed feature
matrix as a projection-method sensitivity check. The PaCMAP result is used to
assess whether the principal qualitative organization persists under a
different nonlinear projection; no PaCMAP-derived score is used as a primary
reported result. Because both methods are stochastic, their random seeds are
fixed in the frozen analysis workflow.

\subsection{Neighborhood label agreement}
\label{sec:supp:purity}

Let \(y_i\) denote the rule-derived label of mode \(i\), and let
\(\mathbf{z}_i\in\mathbb{R}^{2}\) denote its location in a particular
two-dimensional embedding. Define \(\mathcal{N}_k(i)\) as the \(k\) nearest
\emph{other} embedded points to mode \(i\), using Euclidean distance in the
embedding. The focal point is excluded explicitly. In an implementation that
queries nearest neighbors on the fitted sample, at least \(k+1\) candidates
are requested, the index of \(i\) is removed, and the first \(k\) remaining
neighbors are retained.

The per-point same-label agreement is
\begin{equation}
P_i(k)=
\frac{1}{k}
\sum_{j\in\mathcal{N}_k(i)}
\mathbf{1}\!\left\{y_j=y_i\right\},
\label{eq:supp:purity}
\end{equation}
where \(\mathbf{1}\{\cdot\}\) is the indicator function, equal to \(1\) when
its argument is true and \(0\) otherwise. For class \(c\), let
\(\mathcal{I}_c=\{i:y_i=c\}\). The class-level statistic is
\begin{equation}
P_c(k)=
\frac{1}{\lvert\mathcal{I}_c\rvert}
\sum_{i\in\mathcal{I}_c}P_i(k).
\label{eq:supp:class_purity}
\end{equation}

\begin{table}[htb!]
\centering
\small
\caption{Same-label neighborhood agreement in the representative UMAP
embedding for \(k=30\), with the focal point excluded from its neighbor set.
Higher values indicate that a mode is more frequently surrounded by modes
carrying its own rule-derived label.}
\label{tab:supp:purity}

\begin{tabular}{@{}lr@{}}
\toprule
\textbf{Rule-derived class} & \(\boldsymbol{P_c(30)}\) \\
\midrule
ETG & \(0.858\) \\
MTM & \(0.825\) \\
ITG/TEM & \(0.763\) \\
\texttt{Unknown} & \(0.664\) \\
KBM & \(0.644\) \\
\midrule
Macro-average & \(0.752\) \\
\bottomrule
\end{tabular}
\end{table}

Table~\ref{tab:supp:purity} reports the corrected same-label agreement for the
representative UMAP embedding at \(k=30\). The mean is the unweighted
macro-average over the five rule-derived classes. The reported macro-average is calculated from the unrounded
class-specific values.  The ordering is unchanged from that obtained with the earlier
dominant-neighborhood-label statistic, but the interpretation is now
unambiguous: a neighborhood contributes positively only when its label matches
that of the focal mode. In particular, a KBM mode surrounded entirely by
ITG/TEM modes receives \(P_i(k)=0\), not \(1\). At \(k=30\), \(968\) of the
\(6{,}053\) modes (\(16.0\%\)) lie in neighborhoods whose most frequent label
differs from the label of the focal mode. The largest reductions relative to
dominant-label homogeneity occur for KBM and \texttt{Unknown}, consistent with
their stronger embedding within the broader ion-scale population.

These values quantify local consistency of the rule-derived taxonomy in one
selected embedding. They are not classification accuracies, posterior
probabilities, or evidence that any particular agreement threshold has
universal physical significance.

\subsection{Projection-independent class distances}
\label{sec:supp:class_distances}

Neighborhood agreement depends on a nonlinear two-dimensional projection. To
obtain a complementary projection-independent diagnostic, class separations
are computed directly from the transformed vectors
\(\mathbf{x}_i\in\mathbb{R}^{38}\).

For class \(c\), let \(\mathcal{I}_c=\{i:y_i=c\}\). For two distinct classes,
define
\begin{equation}
\mathcal{P}_{ab}
=
\left\{(i,j):i\in\mathcal{I}_a,\,
j\in\mathcal{I}_b\right\},
\qquad a\ne b.
\end{equation}
For a within-class comparison, define
\begin{equation}
\mathcal{P}_{aa}
=
\left\{(i,j):i,j\in\mathcal{I}_a,\ i<j\right\},
\end{equation}
so that self-pairs and duplicated orderings are excluded. The mean class-pair
distance is
\begin{equation}
\overline{d}_{ab}
=
\frac{1}{\lvert\mathcal{P}_{ab}\rvert}
\sum_{(i,j)\in\mathcal{P}_{ab}}
\left\lVert\mathbf{x}_i-\mathbf{x}_j\right\rVert_2.
\label{eq:supp:class_distance}
\end{equation}

Let
\begin{equation}
\mathcal{P}_{\mathrm{all}}
=
\left\{(i,j):1\leq i<j\leq N\right\}
\end{equation}
denote the set of all distinct unordered mode pairs in the cohort, and let
\(\mu_{\mathrm{all}}\) and \(\sigma_{\mathrm{all}}\) be the mean and
population standard deviation of their Euclidean distances. The standardized
class-pair value displayed in Fig.~\ref{fig:standard_scalar_dist} is
\begin{equation}
z_{ab}
=
\frac{\overline{d}_{ab}-\mu_{\mathrm{all}}}
{\sigma_{\mathrm{all}}}.
\label{eq:supp:class_distance_z}
\end{equation}

Negative values therefore denote class-pair separations smaller than the
cohort-wide average, whereas positive values denote larger-than-average
separations. The quantity \(z_{ab}\) is a descriptive standardization, not a
hypothesis-test statistic, and no Gaussian assumption is made for the
pair-distance distribution. This calculation is performed entirely in the
transformed 38-dimensional fingerprint space and is independent of both UMAP
and PaCMAP.

\subsection{Auxiliary consistency and dimension diagnostics}
\label{sec:supp:fingerprint_auxiliary}

A cross-validated one-vs-rest random forest
\cite{breiman2001randomforests} tests whether the rule-derived labels can be
recovered from the transformed fingerprint coordinates. Class-wise areas
under the receiver-operating-characteristic curve are \(0.999\) for ETG,
\(0.997\) for MTM, \(0.991\) for KBM, \(0.986\) for
\texttt{Unknown}, and \(0.978\) for ITG/TEM. The corresponding five-class
predictions show that the principal residual confusion is between ITG/TEM and
KBM. Because the predictors and labels share their physical construction,
these values measure internal recoverability rather than independent
classification performance.

The TwoNN intrinsic-dimension estimator
\cite{facco2017twonn}, applied to the transformed 38-dimensional coordinates,
gives an effective dimension of approximately \(3.5\). Principal-component
analysis provides a complementary linear measure: four principal components
capture \(90\%\) of the variance and five capture \(95\%\).

Finally, a three-component Gaussian-mixture analysis of the \(1{,}126\) MTM
modes produces groups containing \(364\), \(407\), and \(355\) modes. The
bootstrap-stability adjusted Rand index is \(0.83\). The groups differ
principally in electromagnetic-to-electrostatic flux ratios, real frequency,
and mode width. These groups are candidates for subsequent physical study and
should not be interpreted as established MTM subtypes.

Taken together, these diagnostics describe the internal geometry of the
selected fingerprint representation and the reproducibility of a
rule-derived taxonomy. They do not establish that the labels are uniquely
defined physical classes, nor do they demonstrate transfer to other codes,
database snapshots, or independently labeled mode populations.

\section{Parameter-space coverage analysis}
\label{sec:supp:coverage}

The coverage analysis asks how the archived linear simulations occupy a common
input-coordinate system. The frozen cohort is drawn from the
\texttt{LinearRuns} collection of the April~26, 2026 database snapshot.
Of the \(26{,}953\) linear GENE and CGYRO records initially queried,
\(2{,}390\) electron-only calculations are excluded because \(T_i/T_e\) is
undefined. The retained cohort therefore contains \(N=24{,}563\) records
providing all 13 control parameters used below.

\subsection{Normalized control space}
\label{sec:supp:coverage_normalization}

The retained coordinates are
\[
\beta,\quad
q,\quad
\widehat{s},\quad
\alpha_{\mathrm{MHD}},\quad
T_i/T_e,\quad
k_y\rho_s,\quad
R/L_{T_i},\quad
R/L_{T_e},\quad
R/L_{n_i},\quad
R/L_{n_e},\quad
\nu_{ei},\quad
n_z/n_e,\quad
Z.
\]
For parameter \(j\), the normalized coordinate of record \(i\) is
\begin{equation}
x_{ij}
=
\frac{q_{ij}-q_j^{\min}}
     {q_j^{\max}-q_j^{\min}},
\label{eq:supp:minmax}
\end{equation}
where \(q_j^{\min}\) and \(q_j^{\max}\) are computed over the frozen coverage
cohort. Every retained parameter has a nonzero empirical range. Equation
\eqref{eq:supp:minmax} maps the empirical bounding box to
\([0,1]^{13}\) and prevents the numerical scale of any one coordinate from
dominating Euclidean distances.

This normalization does not imply that the coordinates are statistically
independent or that every point in the resulting hypercube represents a
physically realizable plasma. In particular, the uniformly sampled probe
points used below explore the empirical bounding box, including combinations
that may violate physical or campaign-specific correlations. The resulting
global-coverage statistic therefore measures how completely MGKDB fills that
bounding box; it is not a measure of coverage under a physically admissible
prior distribution.

An equally sized Halton low-discrepancy design
\cite{halton1960}, containing \(N=24{,}563\) points, provides an approximately
uniform space-filling reference in the same normalized hypercube. The Halton
design is a geometric baseline rather than a physical prior, an optimal
simulation design, or a model of how gyrokinetic calculations should be
distributed.

\subsection{Principal-component and effective-dimension diagnostics}
\label{sec:supp:coverage_pca}

Principal-component analysis is applied to the centered normalized coordinate
matrix. Let
\[
\lambda_1\geq\lambda_2\geq\cdots\geq\lambda_{13}\geq 0
\]
denote the eigenvalues of its empirical covariance matrix. The cumulative
explained-variance fraction of the first \(m\) components is
\begin{equation}
V_m
=
\frac{\sum_{\ell=1}^{m}\lambda_\ell}
     {\sum_{\ell=1}^{13}\lambda_\ell}.
\label{eq:supp:coverage_variance}
\end{equation}
For the frozen cohort, \(V_4\simeq0.80\) and \(V_5\simeq0.87\).

A complementary effective dimension is obtained from the participation ratio
of the PCA spectrum,
\begin{equation}
d_{\mathrm{PR}}
=
\frac{\left(\sum_{\ell=1}^{13}\lambda_\ell\right)^2}
     {\sum_{\ell=1}^{13}\lambda_\ell^2}.
\label{eq:supp:participation_ratio}
\end{equation}
The coverage cohort gives \(d_{\mathrm{PR}}\simeq4.5\), indicating that the
variance is concentrated along substantially fewer than 13 independent
directions.

Pearson correlations provide a direct view of some of this dependence. The
strongest reported relationships include
\[
r\!\left(R/L_{n_e},R/L_{n_i}\right)=0.98,
\qquad
r\!\left(R/L_{T_e},R/L_{n_e}\right)=0.97.
\]
These correlations reflect the profile relationships and scan designs
represented in the archive; they should not be interpreted as universal
physical constraints.

\subsection{Global and local coverage statistics}
\label{sec:supp:coverage_statistics}

Let
\(\mathbf{x}_i\in[0,1]^{13}\) denote the normalized coordinate vector of
archive entry \(i\). A common set of \(M=5{,}000\) probe points
\(\mathbf{u}_m\) is sampled uniformly from \([0,1]^{13}\). Global coverage at
probe point \(m\) is summarized by the distance to the nearest sample,
\begin{equation}
r^\star(\mathbf{u}_m)
=
\min_{1\leq i\leq N}
\left\lVert
\mathbf{u}_m-\mathbf{x}_i
\right\rVert_2.
\label{eq:supp:global_coverage}
\end{equation}
The same probe points are evaluated against MGKDB and the equally sized Halton
reference. This paired construction ensures that differences between the two
distance distributions do not arise from different probe locations. The
median value is approximately \(1.28\) for MGKDB and \(0.51\) for the Halton
reference.

Local concentration is measured by the mean distance from each entry to its
five nearest \emph{other} entries,
\begin{equation}
\overline{d}_5(i)
=
\frac{1}{5}
\sum_{j\in\mathcal{N}_5(i)}
\left\lVert
\mathbf{x}_i-\mathbf{x}_j
\right\rVert_2,
\label{eq:supp:local_spacing}
\end{equation}
where \(i\notin\mathcal{N}_5(i)\). For positive distances, the median is
approximately \(1.09\times10^{-3}\) for MGKDB and \(0.57\) for the Halton
reference. The \(586\) MGKDB entries with
\(\overline{d}_5(i)=0\) are excluded from the logarithmic histogram and
reported separately.

The global and local statistics answer different questions. The distribution
of \(r^\star\) measures how far arbitrary bounding-box locations lie from the
archive, whereas \(\overline{d}_5\) measures concentration around locations
that the archive actually occupies. A dataset can therefore be globally
sparse while simultaneously containing tightly repeated local scan
structures.

\subsection{Repeated parameter vectors}
\label{sec:supp:coverage_duplicates}

Repetition is evaluated using exact equality of the 13 extracted parameter
values, without additional rounding. With all 13 coordinates retained, the
number of distinct vectors is \(22{,}782\). The number of rows beyond the
first occurrence of each distinct vector is therefore
\begin{equation}
N_{\mathrm{repeat}}
=
N-N_{\mathrm{unique}}
=
24{,}563-22{,}782
=
1{,}781.
\label{eq:supp:duplicate_count}
\end{equation}
Thus, \(1{,}781\) rows are exact copies of an earlier 13-parameter row. This
quantity differs from the number of rows belonging to duplicated groups:
counting every member of every group with multiplicity greater than one gives
\(2{,}754\) rows.

The zero-distance nearest-neighbor statistic is stricter. A record has
\(\overline{d}_5(i)=0\) only when at least five other records have the same
13-parameter vector. The \(586\) zero-distance entries therefore belong to
coincident groups containing at least six rows.

The role of wavenumber scans can be isolated by removing \(k_y\rho_s\) from
the comparison. In the remaining 12 coordinates, the \(24{,}563\) rows reduce
to \(4{,}101\) distinct configurations, corresponding to approximately six
wavenumber-resolved entries per configuration on average. This reduction is
consistent with the prevalence of \(k_y\) scans in the archived campaigns.

These counts establish repetition only within the selected parameter
projection. They do not by themselves demonstrate duplication of complete
simulation records: calculations with identical values in the retained
coordinates can differ in magnetic geometry, species detail, numerical
resolution, collision model, code version, or other fields not included in
the 13-parameter representation.

\subsection{Campaign-planning score}
\label{sec:supp:coverage_ranking}

The parameter-ranking diagnostic combines two descriptive factors: the
contribution of a parameter to the leading PCA subspace and the unevenness of
its marginal sampling. Let \(v_\ell\) be the explained-variance fraction of
principal component \(\ell\), let \(w_{j\ell}\) be the corresponding loading
of parameter \(j\), and let \(\mathcal{L}\) denote the set of leading
components retained by the analysis. The variance-weighted loading score is
\begin{equation}
L_j
=
\frac{
\displaystyle
\sum_{\ell\in\mathcal{L}}
v_\ell w_{j\ell}^2
}{
\displaystyle
\sum_{\ell\in\mathcal{L}}v_\ell
}.
\label{eq:supp:loading_score}
\end{equation}

For a marginal histogram containing \(B\) equal-width bins, let
\(p_{jb}\) be the fraction of parameter-\(j\) observations in bin \(b\).
Its Shannon entropy is
\begin{equation}
H_j
=
-\sum_{b=1}^{B}p_{jb}\log p_{jb},
\qquad
0\log 0\equiv0.
\label{eq:supp:marginal_entropy}
\end{equation}
The normalized under-sampling factor and final score are
\begin{equation}
U_j
=
1-\frac{H_j}{\log B},
\qquad
S_j=L_jU_j.
\label{eq:supp:coverage_score}
\end{equation}
A parameter receives a large score only when it contributes appreciably to the
leading populated subspace and its marginal distribution is uneven.

The highest-ranked coordinate is impurity charge \(Z\), followed by the four
gradient drives
\(R/L_{n_e}\), \(R/L_{n_i}\), \(R/L_{T_e}\), and \(R/L_{T_i}\).
The interpretation of \(Z\) requires particular care: its archived values are
effectively categorical,
\(Z\in\{0,4,6\}\), and carbon accounts for approximately \(94\%\) of the
entries. Its high score therefore signals limited impurity-species diversity,
not a smooth interval that should be filled by interpolating additional
charge values.

The score \(S_j\) is a heuristic archive diagnostic. It does not account for
the computational cost of new simulations, physical feasibility,
measurement uncertainty, target observables, or the expected information gain
from a proposed design. It should therefore be used to identify candidate
directions for further examination, not as an independently sufficient
optimal-experimental-design criterion.

\section{CGYRO--TGLF surrogate workflow}
\label{sec:supp:surrogate}

This demonstration evaluates whether MGKDB can convert archived simulation
records into a traceable dataset for downstream modeling. The source operating
points come from nonlinear CGYRO calculations, but all surrogate targets are
generated by TGLF. The resulting model is therefore an emulator of TGLF
outputs over a parameter set inherited from a CGYRO campaign; it is not a
surrogate for nonlinear CGYRO transport and does not, by itself, assess
TGLF--CGYRO fidelity.

\subsection{Dataset and model construction}
\label{sec:supp:surrogate_model}

The frozen cohort contains \(N=50\) nonlinear CGYRO operating points from one
ITER-startup optimization campaign using the PORTALS workflow
\cite{rodriguezfernandez2024portals}. The cohort comprises ten optimizer iterations
at each of five radial locations,
\begin{equation}
\rho\in\{0.35,0.55,0.75,0.88,0.90\}.
\label{eq:supp:surrogate_radii}
\end{equation}
Candidate records are identified through metadata filters including
\(\texttt{sim\_type = CGYRO}\) and
\(\texttt{IsLinear = False}\). The resolved MGKDB record identifiers are
retained in the frozen cohort manifest.

For each record, the native \texttt{input.cgyro} file is retrieved and
converted into a TGLF input using
\textit{Pyrokinetics}~\cite{patel2024pyrokinetics}. TGLF
\cite{tglf} is then executed at the corresponding operating point. The
software versions, conversion settings, TGLF configuration, and source-record
identifiers form part of the archived workflow provenance.

The resulting input matrix
\(\mathbf{X}\in\mathbb{R}^{50\times38}\) contains 38 scalar features spanning
Miller-geometry coefficients, species densities and temperatures, density and
temperature gradients, \(\beta_e\), collisionality,
\(Z_{\mathrm{eff}}\), and rotation-related quantities. The feature order is
fixed in the archived feature manifest.

The TGLF outputs are organized into targets of increasing dimension. The
principal targets discussed in the manuscript are:

\begin{itemize}
\item \texttt{flux\_energy\_total}: two components containing the electron
energy flux \(Q_e\) and the summed-ion energy flux \(Q_i\);

\item \texttt{flux\_energy}: the corresponding per-species energy-flux
components;

\item \texttt{gamma\_peak}: the three-component vector
\begin{equation}
\mathbf{y}_{\mathrm{peak}}
=
\left(
\gamma_{\max},
k_y^{\max},
\omega_{\max}
\right),
\label{eq:supp:gamma_peak}
\end{equation}
where \(\gamma_{\max}=\gamma(k_y^{\max})\) and
\(\omega_{\max}=\omega(k_y^{\max})\);

\item \texttt{gamma\_ky}: the 33-component growth-rate spectrum
\(\gamma(k_y)\); and

\item \texttt{eigenvalue\_raw}: the 132-component unaggregated eigenvalue
target.
\end{itemize}

The target-component ordering in Eq.~\eqref{eq:supp:gamma_peak} is important:
component \(0\) is \(\gamma_{\max}\), component \(1\) is \(k_y^{\max}\), and
component \(2\) is \(\omega_{\max}\). Panel~(b) of
Fig.~\ref{fig:surrogate_parity} displays and scores only component \(0\). For a target containing \(p\) output components, the primary estimator is the
pipeline
\begin{equation}
\operatorname{StandardScaler}
\;\longrightarrow\;
\operatorname{Ridge}
\!\left(
\alpha=\max\{1,p/2\}
\right).
\label{eq:supp:ridge_pipeline}
\end{equation}
Within each training fold, the feature scaler subtracts the training-fold mean
and divides by the training-fold standard deviation. The fitted scaler is then
applied unchanged to the held-out fold. Target values are not standardized.

For a standardized training matrix \(\mathbf{X}_f\) and target matrix
\(\mathbf{Y}_f\), ridge regression minimizes
\begin{equation}
\left\lVert
\mathbf{Y}_f-\mathbf{X}_f\mathbf{W}
-\mathbf{1}\mathbf{b}^{\mathsf{T}}
\right\rVert_F^2
+
\alpha\left\lVert\mathbf{W}\right\rVert_F^2,
\label{eq:supp:ridge_objective}
\end{equation}
where \(\mathbf{W}\) contains the regression coefficients and
\(\mathbf{b}\) contains the intercepts. The value of \(\alpha\) is fixed by
Eq.~\eqref{eq:supp:ridge_pipeline}; it is not selected using the held-out folds. Five-fold cross-validation uses shuffled folds with random seed \(0\). Each
fold therefore contains 40 training points and 10 held-out points. The same
fold assignments are used across targets and model classes. A fresh copy of
the complete preprocessing-and-regression pipeline is fitted in every fold. A multilayer perceptron provides a nonlinear comparison. It uses hidden layers
of 64 and 32 units, rectified-linear activations, a maximum of \(5{,}000\)
iterations, early stopping with an internal validation fraction of \(0.15\),
and random seed \(0\). This comparison is illustrative: with only 40
training cases per outer fold, it is not an architecture search or a
well-powered comparison between linear and nonlinear model families.

\subsection{Multi-output scoring and parity plots}
\label{sec:supp:surrogate_scoring}

Let \(f\) denote a held-out fold and \(c\) an output component. The
component-specific coefficient of determination is
\begin{equation}
R^2_{f,c}
=
1-
\frac{
\displaystyle
\sum_{i\in f}
\left(y_{ic}-\widehat{y}_{ic}\right)^2
}{
\displaystyle
\sum_{i\in f}
\left(y_{ic}-\overline{y}_{f,c}\right)^2
}.
\label{eq:supp:r2_component}
\end{equation}
Here, \(y_{ic}\) is the observed value of output component \(c\) for
sample \(i\), \(\widehat{y}_{ic}\) is its out-of-fold prediction from a
model trained without fold \(f\), and
\begin{equation}
\overline{y}_{f,c}
=
\frac{1}{\lvert f\rvert}
\sum_{i\in f} y_{ic}
\end{equation}
is the mean observed value of component \(c\) within the held-out fold.
No held-out fold is constant for either
\texttt{flux\_energy\_total} or \texttt{gamma\_peak}, so the denominator is
nonzero for every component used in Fig.~\ref{fig:surrogate_parity}.

For a target with \(p\) components, the primary multi-output fold score is the
uniform average
\begin{equation}
R^2_f
=
\frac{1}{p}
\sum_{c=1}^{p}R^2_{f,c}.
\label{eq:supp:r2_uniform}
\end{equation}
This gives every output component equal weight, irrespective of its numerical
variance. It differs from pooling the residual and total sums of squares
across components, which would weight components according to their
fold-specific variance.

The reported value and fold-to-fold variation are
\begin{equation}
\overline{R^2}
=
\frac{1}{5}\sum_{f=1}^{5}R^2_f,
\qquad
\sigma_{R^2}
=
\left[
\frac{1}{5}
\sum_{f=1}^{5}
\left(R^2_f-\overline{R^2}\right)^2
\right]^{1/2}.
\label{eq:supp:r2_summary}
\end{equation}
Thus, the reported uncertainty is the population standard deviation of the
five foldwise scores (\(\texttt{ddof}=0\)); it is not a confidence interval
or an estimate of uncertainty under repeated campaign sampling.

The scores relevant to Fig.~\ref{fig:surrogate_parity} are summarized in
Table~\ref{tab:supp:surrogate_scores}.

\begin{table}[htb!]
\centering
\small
\caption{Corrected ridge-regression scores for the principal surrogate
targets. Values are means and population standard deviations across the five
held-out folds.}
\label{tab:supp:surrogate_scores}

\begin{tabularx}{\linewidth}{@{}
    >{\raggedright\arraybackslash}X
    >{\raggedright\arraybackslash}p{0.26\linewidth}
    >{\raggedleft\arraybackslash}p{0.19\linewidth}@{}}
\toprule
\textbf{Quantity} &
\textbf{Score reported} &
\(\boldsymbol{\overline{R^2}\pm\sigma_{R^2}}\) \\
\midrule

Electron and summed-ion energy fluxes,
Fig.~\ref{fig:surrogate_parity}(a) &
Two-component uniform average &
\(0.794\pm0.085\) \\

Peak growth rate \(\gamma_{\max}\),
Fig.~\ref{fig:surrogate_parity}(b) &
Component \(0\) of \texttt{gamma\_peak} &
\(0.776\pm0.066\) \\

Complete \texttt{gamma\_peak} target &
Three-component uniform average &
\(0.640\pm0.075\) \\

\bottomrule
\end{tabularx}
\end{table}

Panel~(a) contains out-of-fold predictions for both displayed flux components
and reports the uniform-average score across those two components.
Panel~(b) contains out-of-fold predictions only for
\(\gamma_{\max}\) and therefore reports the component-specific score
\(R^2_{f,0}\), averaged across folds. The complete three-component
\texttt{gamma\_peak} score must not be used to annotate the
\(\gamma_{\max}\)-only panel.

Across the broader target hierarchy, the per-species energy-flux target
attains approximately \(R^2=0.79\pm0.08\), the 33-component
\(\gamma(k_y)\) target attains approximately
\(R^2=0.56\pm0.08\), and raw high-dimensional targets, including the
132-component eigenvalue representation, yield negative cross-validated
\(R^2\) at this sample size. The multilayer perceptron underperforms ridge
regression under the same corrected uniform-average scoring convention. These
comparisons are descriptive of the frozen workflow and should not be
interpreted as a general ranking of surrogate architectures.

\subsection{Scope of the validation}
\label{sec:supp:surrogate_scope}

Every plotted prediction is out of fold, and preprocessing is fitted using
only the corresponding training fold. Nevertheless, shuffled pointwise folds
can place radial locations and optimizer iterations from the same campaign in
both training and test sets. The resulting scores quantify interpolation and
learnability within this 50-case campaign; they do not establish transfer to
an unseen optimizer iteration, radial location, campaign, or device.

More stringent validation would include:

\begin{enumerate}
\item leave-one-optimizer-iteration-out validation, holding out all five
radial locations from one optimization step;

\item leave-one-radius-out validation, holding out one radial location across
all ten optimizer iterations; and

\item leave-one-campaign-out validation once multiple sufficiently comparable
campaigns are available.
\end{enumerate}

Finally, the surrogate emulates TGLF outputs, not nonlinear CGYRO transport.
Testing reduced-model fidelity requires a separately defined comparison
between physically commensurate TGLF and CGYRO observables. The demonstrated
result is therefore the reproducible construction of a cross-fidelity
workflow and an internally coherent TGLF training set, rather than validation
of TGLF or a production-ready transport surrogate.

\section{MGKDB record-schema reference}
\label{sec:supp:schema_reference}

This section provides an implementation-level reference for the principal
fields and artifacts supporting the MGKDB record architecture described in
Sec.~\ref{sec:schema}. It documents the schema used by the current ingestion
workflow and should not be interpreted as an exhaustive specification of every
historical record. Legacy documents may omit fields introduced by later
schema or translation versions, while individual codes and campaigns may
supply additional optional content.

Field names and capitalization are literal and case-sensitive. In the metadata
tables below, a group and field should be combined with the top-level branch
name to form the complete query path. For example,
\texttt{CodeTag} and \texttt{sim\_type} correspond to
\[
\texttt{Metadata.CodeTag.sim\_type}.
\]
Queries should test explicitly for field presence. An absent field must not be
interpreted automatically as a false Boolean value, a numerical zero, or an
empty physical quantity. Reproducible analyses should also record the MGKDB
software version, schema version, and translation version associated with the
queried records whenever these are available.

\subsection{Metadata fields}
\label{sec:supp:metadata_fields}

The \texttt{Metadata} branch combines provenance and grouping information with
scenario, publication, code, and workflow metadata.
Table~\ref{tab:supp:metadata_provenance} lists the principal provenance fields,
and Table~\ref{tab:supp:metadata_scientific} lists the principal scenario and
code fields.

\begin{table}[htb!]
\centering
\small
\caption{Principal provenance and grouping fields under
\texttt{Metadata}. Complete query paths have the form
\texttt{Metadata.DBtag.<field>}. Fields may be absent from legacy records.}
\label{tab:supp:metadata_provenance}

\begingroup
\setlength{\tabcolsep}{4pt}
\renewcommand{\arraystretch}{1.14}

\begin{tabularx}{\linewidth}{@{}
    >{\raggedright\arraybackslash}p{0.16\linewidth}
    >{\raggedright\arraybackslash}p{0.27\linewidth}
    >{\raggedright\arraybackslash}X@{}}
\toprule
\textbf{Group} & \textbf{Field} & \textbf{Interpretation} \\
\midrule

\texttt{DBtag} &
\texttt{user} &
Identifier of the uploader; not necessarily the sole simulation author or the
person who executed the calculation. \\

\texttt{DBtag} &
\texttt{run\_collection\_name} &
Folder-derived grouping path used to organize related uploads; not a globally
persistent campaign identifier. \\

\texttt{DBtag} &
\texttt{run\_suffix} &
Run-level suffix, for example \texttt{\_0001} or \texttt{\_rho.95}. \\

\texttt{DBtag} &
\texttt{keywords} &
User-supplied search tags. \\

\texttt{DBtag} &
\texttt{comments} &
Free-text scientific or operational context, including checks that are not
represented in structured fields. \\

\texttt{DBtag} &
\texttt{confidence} &
Uploader-assigned review-status level from 0 to 3; see
Table~\ref{tab:supp:confidence}. \\

\texttt{DBtag} &
\texttt{time\_uploaded} &
Timestamp of initial ingestion. \\

\texttt{DBtag} &
\texttt{last\_updated} &
Timestamp of the most recent recorded update. \\

\texttt{DBtag} &
\texttt{linked\_objectID} &
Identifier of a related MGKDB record, for example a matched reduced-model and
higher-fidelity calculation. \\

\texttt{DBtag} &
\texttt{archive\_location} &
Location of an external or upstream archive, when supplied. \\

\texttt{DBtag} &
\texttt{IsRestart} &
Restart indicator and, where available, the source record and original
timestep. \\

\bottomrule
\end{tabularx}
\endgroup
\end{table}

\begin{table}[htb!]
\centering
\small
\caption{Principal scenario, publication, code, and workflow fields under
\texttt{Metadata}. Complete paths have the form
\texttt{Metadata.<group>.<field>}. Availability varies across records and
model families.}
\label{tab:supp:metadata_scientific}

\begingroup
\setlength{\tabcolsep}{4pt}
\renewcommand{\arraystretch}{1.14}

\begin{tabularx}{\linewidth}{@{}
    >{\raggedright\arraybackslash}p{0.19\linewidth}
    >{\raggedright\arraybackslash}p{0.25\linewidth}
    >{\raggedright\arraybackslash}X@{}}
\toprule
\textbf{Group} & \textbf{Field} & \textbf{Interpretation} \\
\midrule

\texttt{Scenario\_tag} &
\texttt{experiment} &
Associated experimental device or design study, for example DIII-D, ITER, or
SPARC. \\

\texttt{Scenario\_tag} &
\texttt{scenario\_runid} &
Experiment- or campaign-specific run identifier. \\

\texttt{Publications} &
\texttt{doi} &
DOI or DOI list associated with the record. Publication and dataset DOIs
should be distinguished in the accompanying metadata. \\

\texttt{CodeTag} &
\texttt{sim\_type} &
Source code or model identifier, currently including GENE, CGYRO, and TGLF. \\

\texttt{CodeTag} &
\texttt{IsLinear} &
Boolean indicator for a linear calculation. \\

\texttt{CodeTag} &
\texttt{quasi\_linear} &
Boolean indicator for a quasilinear calculation. \\

\texttt{CodeTag} &
\texttt{Has1DFluxes} &
Boolean indicator that one-dimensional flux data are retained. \\

\texttt{CodeTag} &
\texttt{git\_hash} &
Source-code commit identifier, when known. \\

\texttt{CodeTag} &
\texttt{platform} &
Computing platform on which the calculation was executed. \\

\texttt{CodeTag} &
\texttt{execution\_date} &
Date of execution, when supplied. \\

\texttt{CodeTag} &
\texttt{workflow\_type} &
Parent workflow, for example SMARTS or PORTALS. \\

\bottomrule
\end{tabularx}
\endgroup
\end{table}

The collection containing a document should not be used as the sole
description of its physical model or run type. In particular,
\texttt{NonlinRuns} contains nonlinear GENE and CGYRO calculations together
with quasilinear TGLF evaluations. Model and run class should therefore be
determined from fields such as \texttt{sim\_type}, \texttt{IsLinear}, and
\texttt{quasi\_linear}.

\subsection{Uploader-assigned review status}
\label{sec:supp:confidence_levels}

The field \texttt{Metadata.DBtag.confidence} records the level of review
reported by the uploader. It is not a statistical confidence, probability of
correctness, or certificate of physical validity. The levels concern
verification activity and provenance, which are distinct from validation
against experiment, analytic theory, or an independently implemented
code~\cite{oberkampf2004verification}.

\begin{table}[htb!]
\centering
\small
\caption{Interpretation of the uploader-assigned \texttt{confidence} field.
The levels record review activity and should not be interpreted as
probabilities or universal certificates of numerical or physical validity.}
\label{tab:supp:confidence}

\begingroup
\setlength{\tabcolsep}{4pt}
\renewcommand{\arraystretch}{1.14}

\begin{tabularx}{\linewidth}{@{}
    >{\centering\arraybackslash}p{0.08\linewidth}
    >{\raggedright\arraybackslash}p{0.24\linewidth}
    >{\raggedright\arraybackslash}X@{}}
\toprule
\textbf{Level} & \textbf{Review status} & \textbf{Interpretation} \\
\midrule

0 &
Not assessed &
The output has been ingested, but verification activity has not been
documented. This does not mean that the calculation failed. \\

1 &
Plausibility inspected &
Applicable run-level behavior has been inspected. For a nonlinear turbulence
calculation, examples include flux saturation and the absence of obvious
spectral edge accumulation. Applicable criteria depend on model and run type. \\

2 &
Linked to a checked run &
The calculation was derived from or restarted from a more extensively checked
record. The source record and the relevant lineage should be documented
explicitly. \\

3 &
Convergence check documented &
At least one named quantity has been checked for numerical convergence, with
the varied resolution or parameter and the resulting evidence recorded in
metadata. This level does not imply that every output has been verified or
physically validated. \\

\bottomrule
\end{tabularx}
\endgroup
\end{table}

Because the field is uploader assigned, comparisons of
\texttt{confidence} levels across campaigns should account for differences in
the checks performed and the detail recorded in \texttt{comments}. A higher
level does not make otherwise incompatible records scientifically
commensurate.

\subsection{Representative code-native artifacts}
\label{sec:supp:native_artifacts}

The \texttt{Files} branch retains supported code-native inputs and outputs,
either directly within the parent document or through linked GridFS payloads.
The exact contents depend on source code, simulation type, uploader choices,
and record age. Table~\ref{tab:supp:native_files} is therefore representative
rather than exhaustive.

\begin{table}[htb!]
\centering
\small
\caption{Representative code-native artifacts retained under
\texttt{Files}. Presence is not guaranteed for every record, and additional
campaign-specific artifacts may also be attached.}
\label{tab:supp:native_files}

\begingroup
\setlength{\tabcolsep}{4pt}
\renewcommand{\arraystretch}{1.14}

\begin{tabularx}{\linewidth}{@{}
    >{\raggedright\arraybackslash}p{0.11\linewidth}
    >{\raggedright\arraybackslash}p{0.37\linewidth}
    >{\raggedright\arraybackslash}X@{}}
\toprule
\textbf{Code} &
\textbf{Representative artifacts} &
\textbf{Purpose} \\
\midrule

GENE &
\texttt{parameters}, \texttt{autopar}, \texttt{codemods},
\texttt{nrg}, \texttt{omega}, \texttt{magn\_geometry} &
Input specification, parallel configuration, source modifications, reduced
time traces, linear eigenvalues, and magnetic geometry. \\

CGYRO &
\texttt{input.cgyro}, \texttt{input.cgyro.gen},
\texttt{input.gacode}, \texttt{out.cgyro.info} &
Native and generated inputs, optional GACODE-format input, and retained
run-summary output. \\

TGLF &
\texttt{input.tglf}, \texttt{input.tglf.gen},
\texttt{output.tglf.run} &
Native and generated inputs and retained reduced-model output. \\

\bottomrule
\end{tabularx}
\endgroup
\end{table}

The presence of an input file does not by itself guarantee bitwise
reproducibility. Reproduction may additionally require the source-code commit
or executable, dependencies, platform information, external geometry or
equilibrium files, runtime configuration, and other artifacts that are not
complete for every legacy record. The \texttt{Files} branch should therefore
be understood as preserving the strongest available code-native evidence for
each archived run.

\subsection{Standardized and derived branches}
\label{sec:supp:standardized_branches}

For currently supported records, \textit{Pyrokinetics}
\cite{patel2024pyrokinetics} translates available native inputs and outputs
into an Integrated Modelling and Analysis Suite
(IMAS)-aligned~\cite{Imbeaux_2015} representation stored under the
case-sensitive top-level branch \texttt{gyrokinetics}. A group can be absent
when it is inapplicable to the source model, unavailable in the retained
native artifacts, or unsupported by the translation version used for that
record.

\begin{table}[htb!]
\centering
\small
\caption{Principal groups under the IMAS-aligned
\texttt{gyrokinetics} branch. Availability varies with source model,
simulation type, retained artifacts, and translation version.}
\label{tab:supp:imas_groups}

\begingroup
\setlength{\tabcolsep}{4pt}
\renewcommand{\arraystretch}{1.14}

\begin{tabularx}{\linewidth}{@{}
    >{\raggedright\arraybackslash}p{0.31\linewidth}
    >{\raggedright\arraybackslash}X@{}}
\toprule
\textbf{Group} & \textbf{Contents} \\
\midrule

\texttt{ids\_properties} &
Identity, provenance, and version information for the standardized data
structure. \\

\texttt{normalizing\_quantities} &
Reference quantities defining the normalization convention. \\

\texttt{model} &
Model descriptors, including applicable linear or nonlinear, collisional, and
electromagnetic assumptions. \\

\texttt{flux\_surface} &
Local magnetic-geometry quantities, including safety factor, magnetic shear,
elongation, and triangularity when available. \\

\texttt{species\_all} &
Quantities applying across species, including reference beta and effective
charge when available. \\

\texttt{species} &
Per-species density, temperature, gradients, mass, charge, and related
quantities. \\

\texttt{collisions} &
Collision-model and normalization information. \\

\texttt{linear} / \texttt{non\_linear} &
Available linear eigenmode quantities or nonlinear transport and spectral
outputs. \\

\texttt{code} &
Source-code identity and version information when available. \\

\texttt{time} &
Time coordinates associated with time-dependent quantities. \\

\bottomrule
\end{tabularx}
\endgroup
\end{table}

Where populated, \texttt{ids\_properties} should identify the standardized
representation and its version. The standardized branch supports common-field
queries and candidate comparisons, but matching field paths does not by itself
establish physical equivalence. Cross-code or cross-fidelity comparisons must
also account for geometry, species content, normalization, model assumptions,
resolution, convergence, and observable definitions.

The \texttt{Diagnostics} branch contains cached products derived for recurring
analysis and visualization and is currently populated primarily for GENE
records. It can contain field and moment arrays, computational grids, and
eigenvalue arrays. The optional \texttt{Plots} branch contains convenience
visualizations. Both are derived, non-authoritative layers.

Interpretation and reproduction should trace derived quantities to the
retained native artifacts, the standardized representation, and the versions
of the software used to produce each translation or diagnostic. This
distinction allows the standardized and derived layers to evolve without
erasing the model-specific evidence preserved with the original calculation.

\section{Operational interface reference}
\label{sec:supp:interfaces}

This section documents the operational procedures summarized in
Sec.~\ref{sec:interfaces}. It is intended to make the workflow reproducible
without treating the manuscript as a permanent command-line manual.
Hostnames, authentication procedures, software environments, and client
arguments can evolve independently of the scientific schema. The
version-matched documentation and command help distributed with MGKDB
therefore remain authoritative for current operational details. All examples
below use placeholders for deployment-specific hosts, accounts, paths, and
credentials.

\subsection{Access, installation, and provenance}
\label{sec:supp:access}

The MGKDB client is distributed on the Python Package Index under the package
name \texttt{MGKDB-fusion}; its Python import namespace is \texttt{mgkdb}.
The supported package can be installed with

\begin{small}
\begin{verbatim}
python -m pip install MGKDB-fusion
\end{verbatim}
\end{small}

\noindent or through the version-controlled environment specification in the
MGKDB repository. For reproducible work, users should record the installed
package version, the repository commit or release when repository scripts are
used, and the complete post-processing environment. The package and repository
interfaces should not be assumed to be interchangeable across versions.

Access to the production database requires an account at the National Energy
Research Scientific Computing Center (NERSC) and separate authorization for
MGKDB. The current access-request and credential-configuration procedures are
maintained in the repository documentation at
\url{https://github.com/Sophelio/MGKDB}. Credential files and connection
strings contain secrets and must not be committed to a repository, embedded
in notebooks, included in figures, or placed in shared configuration files.
They should be stored outside version-controlled analysis directories and
protected with appropriate local file permissions.

A reproducible extraction should record, at minimum,

\begin{enumerate}
    \item the database extraction date or frozen snapshot identifier;
    \item the source collection;
    \item the complete query and field projection;
    \item the ObjectID of every selected record;
    \item the requested structured fields and native files;
    \item every post-query exclusion and transformation; and
    \item the software versions used for extraction and analysis.
\end{enumerate}

MongoDB ObjectIDs provide persistent record identifiers and therefore connect
derived data products to their originating MGKDB records. An ObjectID is not,
however, a content hash and does not by itself preserve an earlier state of a
record that is subsequently updated. Reproducibility therefore also requires
a frozen extract, dated database snapshot, or archived analysis product with
appropriate checksums.

\subsection{Uploading and updating records}
\label{sec:supp:upload}

Simulation records are contributed through the code-aware ingestion workflow
implemented by \texttt{mgk\_uploader.py}. The uploader recognizes the
code-specific source layouts documented in the repository. In the current
workflow, GENE runs are commonly distinguished by four-digit suffixes within
a shared directory, for example \texttt{parameters\_0001}. Individual CGYRO
and TGLF runs are commonly stored in separate subdirectories beneath a
campaign directory. Minimal examples of the supported layouts are provided
in the repository's \texttt{test\_data} directory.

A minimal invocation from a version-matched repository checkout has the form

\begin{small}
\begin{verbatim}
python mgk_uploader.py -T <target-directory> \
    -A <credential-file>
\end{verbatim}
\end{small}

\noindent where the angle-bracketed placeholders are replaced locally.
Table~\ref{tab:supp:upload_args} summarizes the principal uploader arguments
used by the workflow represented in this manuscript. The output of
\texttt{-{}-help} from the same software version remains the authoritative
description of the available arguments and their behavior.

\begin{table}[htbp]
\centering
\small
\caption{Principal arguments of \texttt{mgk\_uploader.py} for the software
interface represented in this manuscript. These options are operational
interfaces rather than elements of the scientific schema and may change
between software releases.}
\label{tab:supp:upload_args}

\begingroup
\setlength{\tabcolsep}{5pt}
\renewcommand{\arraystretch}{1.15}

\begin{tabularx}{\linewidth}{@{}
    >{\raggedright\arraybackslash}p{0.16\linewidth}
    >{\raggedright\arraybackslash}p{0.20\linewidth}
    >{\raggedright\arraybackslash}X@{}}
\toprule
\textbf{Flag} & \textbf{Default or status} & \textbf{Purpose} \\
\midrule

\texttt{-T} &
Required &
Target directory containing the simulation outputs to ingest. \\

\texttt{-A} &
Prompted if omitted &
Path to the locally stored MGKDB credential file. \\

\texttt{-L} &
False &
Include supported large field, moment, or velocity-space payloads. \\

\texttt{-Ex} &
False &
Include additional user-selected artifacts. \\

\texttt{-X} &
None &
Exclude selected subdirectories from ingestion. \\

\texttt{-K} &
No keywords &
Attach comma-delimited search keywords to the uploaded records. \\

\texttt{-C} &
0 &
Set the uploader-assigned review-status level from 0 to 3, as defined in
Table~\ref{tab:supp:confidence}. \\

\texttt{-lf}/\texttt{-ls} &
None &
Link the new record to another ObjectID supplied through a file or directly
as a string. \\

\texttt{-V} &
False &
Enable verbose reporting. \\

\texttt{-D} &
False &
Accept configured defaults without interactive prompts. \\

\bottomrule
\end{tabularx}
\endgroup
\end{table}

Large field, moment, and velocity-space arrays can dominate both storage and
transfer costs. They should be included when required for the archival or
scientific purpose of a record, rather than enabled automatically for every
ingestion.

Permitted metadata fields can be amended with
\texttt{support/update\_metadata.py}. Typical operations include associating
publication identifiers, appending explanatory comments, and revising
review-status information. Re-ingestion of a recognized directory can update
files shared across that directory or files associated with a particular run
suffix. When scientific content is updated, the corresponding standardized
\texttt{gyrokinetics} representation and applicable derived diagnostics may
also need to be regenerated. Users should inspect the version-matched command
help, uploader log, and resulting database record before treating an update
as complete.

\subsection{Querying and downloading records}
\label{sec:supp:download}

The command-line retrieval utility \texttt{mgk\_download.py} can select a
campaign through the literal metadata field
\texttt{Metadata.DBtag.run\_collection\_name}, retrieve a specific record by
ObjectID, or recover an individual native file through its stored file
identifier. For example, a group of records from the
\texttt{LinearRuns} collection can be retrieved with a command of the form

\begin{small}
\begin{verbatim}
python mgk_download.py -A <credential-file> \
    -C LinearRuns -T <run-collection-name> \
    -D <output-directory>
\end{verbatim}
\end{small}

\noindent The meanings of similarly named flags are command-specific and
should always be checked using the version-matched
\texttt{-{}-help} output.

For structured analyses, Python and PyMongo can be used to construct compound
MongoDB filters and field projections. Projecting only the required fields
reduces transfer volume and makes the numerical cohort easier to audit.
Native artifacts can then be downloaded only for the selected records that
require source-level inspection or reprocessing.

Large native artifacts may be retained using MongoDB GridFS, which stores a
file as linked database objects rather than as a field embedded directly in
the principal record. Consequently, exporting only the document metadata does
not necessarily export all associated native files. Analyses that depend on
these artifacts should retain their file identifiers, filenames, checksums
when available, and the software procedure used to retrieve them.

The exact query, field projection, extraction date, resolved ObjectIDs, and
post-query exclusions should be written to the analysis provenance before
the data are transformed. This practice distinguishes a reproducible
scientific cohort from a live query whose membership can change as new
records are added or existing records are updated.

\subsection{Interactive inspection with MongoDB Compass}
\label{sec:supp:compass}

MongoDB Compass provides a graphical interface for inspecting the same
collections and nested documents accessed through Python. The production
instance can be reached through a Secure Shell (SSH) tunnel established using
the current protected NERSC and MGKDB connection information. A generic tunnel
template is

\begin{small}
\begin{verbatim}
ssh -N -L <local-port>:<database-host>:<database-port> \
    <nersc-user>@perlmutter.nersc.gov
\end{verbatim}
\end{small}

\noindent after which Compass connects to the forwarded local port using the
MGKDB credentials and authentication database specified in the current
documentation. Database hostnames, ports, and authentication settings should
be obtained from the protected connection instructions and should not be
copied from an older manuscript or shared configuration file.

Compass is useful for identifying collections, inspecting nested field paths,
examining representative values and missing fields, and testing candidate
filters. It should be treated primarily as an inspection and query-development
interface. Any filter used to produce a published result should be exported
or re-expressed in a version-controlled script, and the resulting ObjectIDs
should be retained with the analysis products. Users with write-enabled
accounts should not perform ad hoc production edits through the graphical
interface; record updates should use the supported ingestion and
metadata-update workflows.

\subsection{Local staging and controlled production merge}
\label{sec:supp:local_merge}

MGKDB can be operated against a local MongoDB instance when simulations are
generated on a remote computing resource or when a large campaign should be
validated before transfer. Local records should be created through the same
code-aware ingestion path used for production records and inspected for the
expected metadata, native-file associations, standardized content, and
derived diagnostics.

Transfer from a local instance to the NERSC production archive is an
administrator-controlled operation, not a routine end-user command. Authorized
maintainers should follow the version-matched procedure in the MGKDB
repository, presently documented under

\begin{center}
\path{docs_info/merge_localDB_to_NERSC.md}.
\end{center}

\noindent Referring to the repository path rather than a branch-specific URL
reduces the risk that the manuscript will retain a stale documentation link
after the repository structure or default branch changes.

Before and after a production merge, administrators should verify record
counts, ObjectID handling, cross-record links, GridFS associations, schema
versions, and the presence of required standardized branches. They should
also record the source and destination database states, the software version
used for the transfer, and the applicable collision or replacement policy.
Such policies must be taken from the current administrative procedure and
must not be inferred from filenames or directory structure alone.

\clearpage

\section*{Data and code availability}

The MGKDB source code, client tools, and documentation are publicly available
at \url{https://github.com/Sophelio/MGKDB}. The supported Python distribution
is available from PyPI as \texttt{MGKDB-fusion} at
\url{https://pypi.org/project/MGKDB-fusion/}, with the import namespace
\texttt{mgkdb}. The public datasets, analysis scripts, figure-generation scripts, and other
materials supporting this manuscript are available at
\url{https://github.com/Sophelio/MGKDB_paper_package}. The analyses use the
dated, fixed cohorts identified in Table~\ref{tab:snapshot} and described in
the Supplementary Material. The curated production database is hosted at NERSC and is not available for
unrestricted public download. Direct access to the database or extraction of
additional records requires both NERSC authorization and MGKDB credentials,
as described in the MGKDB repository documentation.

\section*{Declaration of competing interest}

The authors declare the following financial interests or personal
relationships that may be considered potential competing interests: the
authors affiliated with Sophelio LLC disclose that Sophelio LLC develops and
maintains the MGKDB software described in this article. The remaining authors
declare that they have no known competing financial interests or personal
relationships that could have appeared to influence the work reported in this
article.

\section*{Acknowledgements}
This research used resources of the National Energy Research Scientific Computing Center (NERSC), a U.S. Department of Energy Office of Science User Facility, under NERSC award FES-ERCAP0029015.  This work was partially supported by the U.S. Department of Energy, Office of Science, Office of Fusion Energy Sciences, using the DIII-D National Fusion Facility, a DOE Office of Science user facility, under Award No. DE-FC02-04ER54698, along with Office of Fusion Energy Sciences Awards No. DE-SC0024426 and No. DE-SC0024399. 

This report was prepared as an account of work sponsored by an agency of the United States Government. Neither the United States Government nor any agency thereof, nor any of their employees, makes any warranty, express or implied, or assumes any legal liability or responsibility for the accuracy, completeness, or usefulness of any information, apparatus, product, or process disclosed, or represents that its use would not infringe privately owned rights. Reference herein to any specific commercial product, process, or service by trade name, trademark, manufacturer, or otherwise does not necessarily constitute or imply its endorsement, recommendation, or favoring by the United States Government or any agency thereof. The views and opinions of authors expressed herein do not necessarily state or reflect those of the United States Government or any agency thereof.

\section*{Declaration of generative AI and AI-assisted technologies
in the manuscript preparation process}

During the preparation of this work, the authors used Claude, ChatGPT,
and xAI Grok to assist with manuscript organization, language editing,
\LaTeX{} formatting, reference checking, and code review. The authors
reviewed and edited all resulting material, independently verified the
scientific claims, citations, analysis code, and numerical results, and
take full responsibility for the content of the published article.


\begin{thebibliography}{99}

\bibitem{gyro1}
E.A. Frieman, L. Chen,
Nonlinear gyrokinetic equations for low-frequency electromagnetic waves in general plasma equilibria,
Phys. Fluids 25 (1982) 502--508.
\url{https://doi.org/10.1063/1.863762}.

\bibitem{gyro2}
A.J. Brizard, T.S. Hahm,
Foundations of nonlinear gyrokinetic theory,
Rev. Mod. Phys. 79 (2007) 421--468.
\url{https://doi.org/10.1103/RevModPhys.79.421}.

\bibitem{gene}
F. Jenko, W. Dorland, M. Kotschenreuther, B.N. Rogers,
Electron temperature gradient driven turbulence,
Phys. Plasmas 7 (2000) 1904--1910.
\url{https://doi.org/10.1063/1.874014}.

\bibitem{cgyro}
J. Candy, E.A. Belli, R.V. Bravenec,
A high-accuracy Eulerian gyrokinetic solver for collisional plasmas,
J. Comput. Phys. 324 (2016) 73--93.
\url{https://doi.org/10.1016/j.jcp.2016.07.039}.

\bibitem{tglf}
G.M. Staebler, J.E. Kinsey, R.E. Waltz,
A theory-based transport model with comprehensive physics,
Phys. Plasmas 14 (2007) 055909.
\url{https://doi.org/10.1063/1.2436852}.

\bibitem{vandeplassche2020}
K.L. van de Plassche, J. Citrin, C. Bourdelle, Y. Camenen,
F.J. Casson, V.I. Dagnelie, F. Felici, A. Ho, S. Van Mulders,
JET Contributors,
Fast modeling of turbulent transport in fusion plasmas using neural networks,
Phys. Plasmas 27 (2020) 022310.
\url{https://doi.org/10.1063/1.5134126}.

\bibitem{rodriguezfernandez2024portals}
P. Rodriguez-Fernandez, N.T. Howard, A. Saltzman, S. Kantamneni,
J. Candy, C. Holland, M. Balandat, S. Ament, A.E. White,
Enhancing predictive capabilities in fusion burning plasmas through
surrogate-based optimization in core transport solvers,
Nucl. Fusion 64 (2024) 076034.
\url{https://doi.org/10.1088/1741-4326/ad4b3d}.

\bibitem{mdsplus}
J.A. Stillerman, T.W. Fredian, K.A. Klare, G. Manduchi,
MDSplus data acquisition system,
Rev. Sci. Instrum. 68 (1997) 939--942.
\url{https://doi.org/10.1063/1.1147719}.

\bibitem{mdsplus2}
T. Fredian, J. Stillerman, G. Manduchi, A. Rigoni, K. Erickson, T. Schr\"oder,
MDSplus yesterday, today and tomorrow,
Fusion Eng. Des. 127 (2018) 106--110.
\url{https://doi.org/10.1016/j.fusengdes.2017.12.010}.

\bibitem{mdsplusml}
J. Stillerman, S. Lane-Walsh, M. Winkel, C. Rea, G.L. Trevisan, A. Jelenak, J. Readey,
MDSplusML---Optimizations for data access to facilitate machine-learning pipelines,
Fusion Eng. Des. 211 (2025) 114770.
\url{https://doi.org/10.1016/j.fusengdes.2024.114770}.

\bibitem{Imbeaux_2015}
F. Imbeaux, S. Pinches, J.B. Lister, et al.,
Design and first applications of the ITER Integrated Modelling and Analysis Suite,
Nucl. Fusion 55 (2015) 123006.
\url{https://doi.org/10.1088/0029-5515/55/12/123006}.

\bibitem{camenen2026gyrokineticsids}
Y. Camenen, F. Imbeaux, G. Fuhr, T. G\"orler, A. Najlaoui,
Design of the IMAS interface data structure for flux-tube gyrokinetic simulations,
Open Plasma Science 2 (2026) 2.
\url{https://doi.org/10.46298/ops.17447}.

\bibitem{iter_imas_open_2025}
ITER Organization,
Release of IMAS infrastructure and physics models as open source,
8 December 2025.
\url{https://www.iter.org/node/20687/release-imas-infrastructure-and-physics-models-open-source}
(accessed 12 August 2026).

\bibitem{patel2024pyrokinetics}
B.S. Patel, P. Hill, L.T. Pattinson, et al.,
Pyrokinetics---A Python library to standardise gyrokinetic analysis,
J. Open Source Softw. 9 (2024) 5866.
\url{https://doi.org/10.21105/joss.05866}.

\bibitem{gkdb}
G. Fuhr, Y. Camenen, K.L. van de Plassche, F. Almuhisen,
C. Bourdelle, J. Citrin, A. Ho, M. Lanzarone, A. Najlaoui,
Gyro-Kinetic DataBase project,
poster presented at the IAEA Workshop on AI for Accelerating Fusion and
Plasma Science, Vienna, Austria, 28 November--1 December 2023.
\url{https://conferences.iaea.org/event/335/contributions/28984/attachments/15722/26540/Poster_IAEA_2023_GF_LR.pdf}.

\bibitem{materialscloud}
L. Talirz, S. Kumbhar, E. Passaro, et al.,
Materials Cloud, a platform for open computational science,
Sci. Data 7 (2020) 299.
\url{https://doi.org/10.1038/s41597-020-00637-5}.

\bibitem{materialsproject}
A. Jain, S.P. Ong, G. Hautier, et al.,
Commentary: The Materials Project: a materials genome approach to accelerating materials innovation,
APL Mater. 1 (2013) 011002.
\url{https://doi.org/10.1063/1.4812323}.

\bibitem{oqmd}
J.E. Saal, S. Kirklin, M. Aykol, B. Meredig, C. Wolverton,
Materials design and discovery with high-throughput density functional theory: the Open Quantum Materials Database (OQMD),
JOM 65 (2013) 1501--1509.
\url{https://doi.org/10.1007/s11837-013-0755-4}.

\bibitem{fair}
M.D. Wilkinson, M. Dumontier, I.J. Aalbersberg, et al.,
The FAIR Guiding Principles for scientific data management and stewardship,
Sci. Data 3 (2016) 160018.
\url{https://doi.org/10.1038/sdata.2016.18}.

\bibitem{dudson2009bout}
B.D. Dudson, M.V. Umansky, X.Q. Xu, P.B. Snyder, H.R. Wilson,
BOUT++: A framework for parallel plasma fluid simulations,
Comput. Phys. Commun. 180 (2009) 1467--1480.
\url{https://doi.org/10.1016/j.cpc.2009.03.008}.

\bibitem{fair4rs}
M. Barker, N.P. Chue Hong, D.S. Katz, A.-L. Lamprecht,
C. Martinez-Ortiz, F. Psomopoulos, J. Harrow, L.J. Castro,
M. Gruenpeter, P.A. Martinez, T. Honeyman,
Introducing the FAIR Principles for research software,
Sci. Data 9 (2022) 622.
\url{https://doi.org/10.1038/s41597-022-01710-x}.

\bibitem{banker2012mongodb}
K. Banker,
MongoDB in Action,
Manning Publications Co., Shelter Island, NY, 2012.
ISBN 978-1-935182-87-0.

\bibitem{chodorow2010mongodb}
K. Chodorow, M. Dirolf,
MongoDB: The Definitive Guide,
1st ed.,
O'Reilly Media, Inc., Sebastopol, CA, 2010.
ISBN 978-1-4493-8156-1.

\bibitem{lyons2023step}
B.C. Lyons, J. McClenaghan, T. Slendebroek, et al.,
Flexible, integrated modeling of tokamak stability, transport, equilibrium,
and pedestal physics,
Phys. Plasmas 30 (2023) 092510.
\url{https://doi.org/10.1063/5.0156877}.

\bibitem{oberkampf2004verification}
W.L. Oberkampf, T.G. Trucano, C. Hirsch,
Verification, validation, and predictive capability in computational
engineering and physics,
Appl. Mech. Rev. 57 (2004) 345--384.
\url{https://doi.org/10.1115/1.1767847}.

\bibitem{hatch_22}
D.R. Hatch, C. Michoski, D. Kuang, et al.,
Reduced models for ETG transport in the tokamak pedestal,
Phys. Plasmas 29 (2022) 062501.
\url{https://doi.org/10.1063/5.0087403}.

\bibitem{Kotschenreuther2019}
M. Kotschenreuther, X. Liu, D.R. Hatch, et al.,
Gyrokinetic analysis and simulation of pedestals to identify the culprits for energy losses using ``fingerprints'',
Nucl. Fusion 59 (2019) 096001.
\url{https://doi.org/10.1088/1741-4326/ab1fa2}.

\bibitem{McInnes2018UMAP}
L. McInnes, J. Healy, J. Melville,
UMAP: Uniform Manifold Approximation and Projection for dimension reduction,
arXiv:1802.03426 (2018).
\url{https://doi.org/10.48550/arXiv.1802.03426}.

\bibitem{Wang2021PaCMAP}
Y. Wang, H. Huang, C. Rudin, Y. Shaposhnik,
Understanding how dimension reduction tools work: an empirical approach to
deciphering t-SNE, UMAP, TriMap, and PaCMAP for data visualization,
J. Mach. Learn. Res. 22 (2021), Article 201, 1--73.
\url{https://jmlr.org/papers/v22/20-1061.html}.

\bibitem{meneghini2024fuse}
O. Meneghini, T. Slendebroek, B.C. Lyons, K. McLaughlin,
J. McClenaghan, L. Stagner, J. Harvey, T.F. Neiser,
A. Ghiozzi, G. Dose, J. Guterl, A. Zalzali, T. Cote,
N. Shi, D. Weisberg, S.P. Smith, B.A. Grierson, J. Candy,
FUSE (Fusion Synthesis Engine): A Next Generation Framework for Integrated
Design of Fusion Pilot Plants,
arXiv:2409.05894 (2024).
\url{https://doi.org/10.48550/arXiv.2409.05894}.

\bibitem{hatch26}
D.R. Hatch, L.A. Leppin, M.T. Kotschenreuther, S. Houshmandyar, S.M. Mahajan, J. Schmidt, P.-Y. Li,
Microtearing thresholds and second-stable ballooning in the DIII-D pedestal: Reduced modeling and core-edge implications,
Phys. Plasmas 33 (2026) 072512.
\url{https://doi.org/10.1063/5.0337081}.

\bibitem{predebon_13}
I. Predebon, F. Sattin,
On the linear stability of collisionless microtearing modes,
Phys. Plasmas 20 (2013) 040701.
\url{https://doi.org/10.1063/1.4799980}.

\bibitem{hassan_22}
E. Hassan, D.R. Hatch, M.R. Halfmoon, M. Curie, M.T. Kotschenreuther, S.M. Mahajan, G. Merlo, R.J. Groebner, A.O. Nelson, A. Diallo,
Identifying the microtearing modes in the pedestal of DIII-D H-modes using gyrokinetic simulations,
Nucl. Fusion 62 (2022) 026008.
\url{https://doi.org/10.1088/1741-4326/ac3be5}.

\bibitem{breiman2001randomforests}
L. Breiman,
Random forests,
Mach. Learn. 45 (2001) 5--32.
\url{https://doi.org/10.1023/A:1010933404324}.

\bibitem{facco2017twonn}
E. Facco, M. d'Errico, A. Rodriguez, A. Laio,
Estimating the intrinsic dimension of datasets by a minimal neighborhood
information,
Sci. Rep. 7 (2017) 12140.
\url{https://doi.org/10.1038/s41598-017-11873-y}.

\bibitem{halton1960}
J.H. Halton,
On the efficiency of certain quasi-random sequences of points in evaluating
multi-dimensional integrals,
Numer. Math. 2 (1960) 84--90.
\url{https://doi.org/10.1007/BF01386213}.

\bibitem{hoerl1970ridge}
A.E. Hoerl, R.W. Kennard,
Ridge regression: biased estimation for nonorthogonal problems,
Technometrics 12 (1970) 55--67.
\url{https://doi.org/10.1080/00401706.1970.10488634}.

\bibitem{Meneghini_2021}
O. Meneghini, G. Snoep, B.C. Lyons, et al.,
Neural-network accelerated coupled core-pedestal simulations with self-consistent transport of impurities and compatible with ITER IMAS,
Nucl. Fusion 61 (2021) 026006.
\url{https://doi.org/10.1088/1741-4326/abb918}.

\end{thebibliography}
\end{document}